\documentclass{article}

\usepackage{arxiv}

\usepackage{cite}
\usepackage{amsmath,amssymb,amsfonts}
\usepackage{algorithmic}
\usepackage{graphicx}
\usepackage{textcomp}
\usepackage{xcolor}
\usepackage{balance}

\usepackage{balance}  
\usepackage{graphicx}
\usepackage{multicol}
\usepackage{xcolor}
\usepackage{hyperref}
\hypersetup{colorlinks=false}
\usepackage{float}
\usepackage{multirow}
\usepackage{subcaption}
\usepackage{amsmath}
\usepackage[inline]{enumitem}
\usepackage{lipsum}
\usepackage{paralist}
\usepackage[colorinlistoftodos]{todonotes}
\usepackage{authblk}
\usepackage{mathtools}

\newcommand{\new}[1]{#1}

\newcommand{\parg}[1]{\vspace{1mm}\textbf{#1.}\hspace{2mm}}

\def\BibTeX{{\rm B\kern-.05em{\sc i\kern-.025em b}\kern-.08em
    T\kern-.1667em\lower.7ex\hbox{E}\kern-.125emX}}
\begin{document}

\title{Siamese Graph Neural Networks for Data Integration}

\author[1, 2]{Evgeny Krivosheev}
\author[1]{Mattia Atzeni}
\author[1]{Katsiaryna Mirylenka}
\author[1]{Paolo Scotton}
\author[2]{Fabio Casati}

\affil[1]{IBM Research -- Zurich, Switzerland\\
{\{kri, atz, kmi, psc\}@zurich.ibm.com}}
\affil[2]{University of Trento, Italy\\
\{evgeny.krivosheev, fabio.casati\}@unitn.it}


\maketitle

\begin{abstract}
Data integration has been studied extensively for decades and approached from different angles. However, this domain still remains largely rule-driven and  lacks universal automation.
Recent development in machine learning and in particular deep learning has opened the way to more general and more efficient solutions to data integration problems. 
In this work, we propose a general approach to modeling and integrating entities from structured data, such as relational databases,
as well as unstructured  sources, such as free text from news articles.
Our approach is designed to explicitly model and leverage relations between entities, thereby using all available information and preserving as much context as possible.
This is achieved by combining siamese and graph neural networks to propagate information between connected entities \new{and support high scalability.}
We evaluate our method on the task of integrating data about business entities, and we demonstrate that it outperforms standard rule-based systems, as well as other deep learning approaches that do not use graph-based representations.
\end{abstract}


\section{Introduction}
\label{sec:introduction}

Although knowledge graphs (KGs) and ontologies have been exploited effectively for data integration~\cite{pershina2015, Trivedi2018, Azmy2019}, entity matching involving structured and unstructured sources has usually been performed by treating records such as arrays or  text without explicitly taking into account the natural graph representation of structured sources and the potential graph representations of unstructured data~\cite{Getoor2012, Gottapu2016, Mudgal2018, Dong2018, gschwind2019}. 
Implicit graph representations of database records have so far been exploited for record-linkage tasks by taking into account the similarity of the attributes between a query and a candidate record.
Similarities between records are usually computed by means of multicriteria scoring without considering the direct attribute for information flow~\cite{gschwind2019, mirylenka2019linking, mirylenka2019similarity}. Additionally,  many contextual fields, that are often represented by record attributes, are considered to be uninformative and hence are dropped in preprocessing steps, even though they might contain important structural information about the entities represented by the records.


\begin{figure}[t!]
	\begin{center}
		\includegraphics[width=0.45\textwidth]{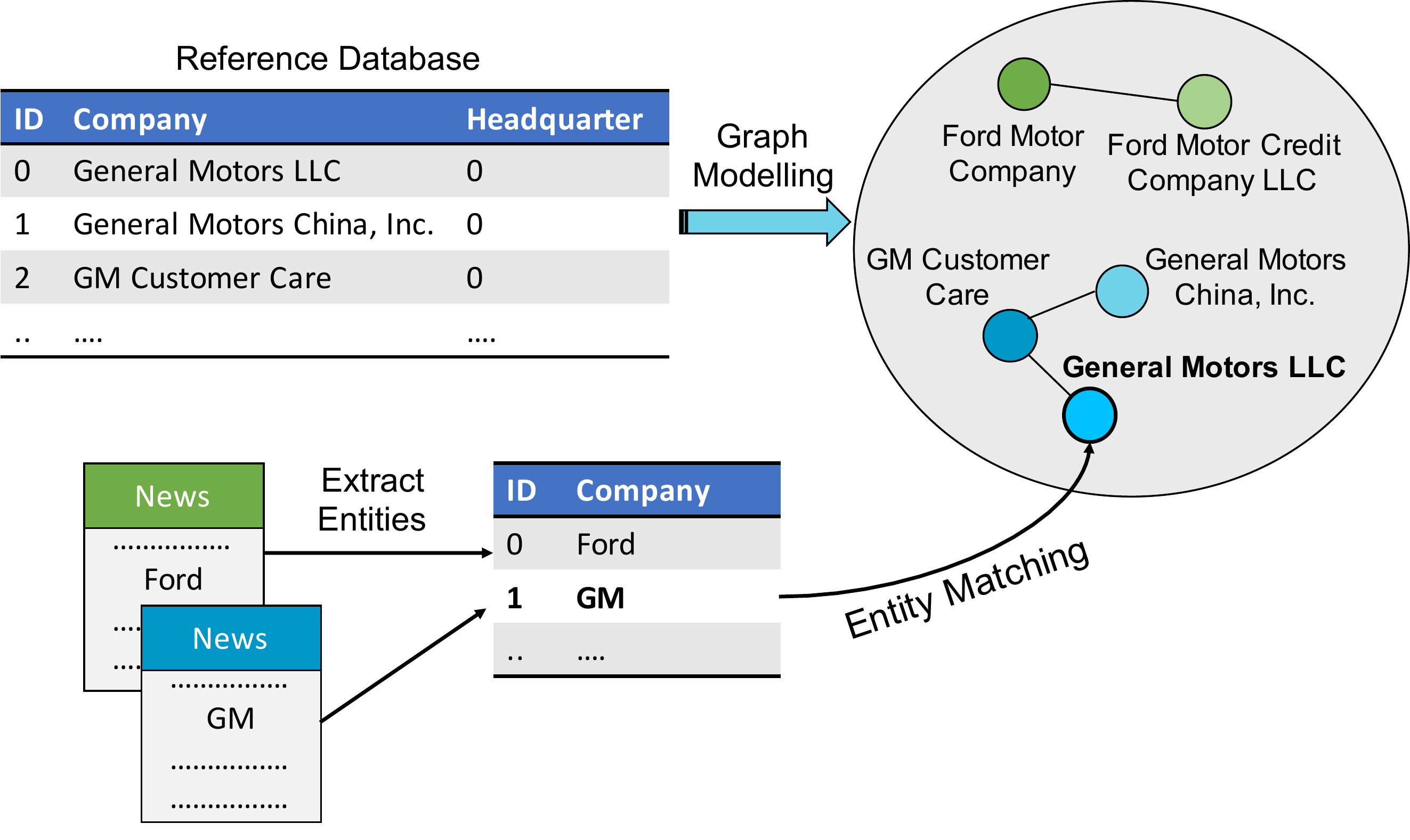}  
		\caption{Example of entity matching and graph modeling}
		\label{fig:linkage-example}
	\end{center}
\end{figure}

To address this issue, this paper introduces a methodology for leveraging graph-structured information in entity matching.
The main idea consists of exploiting KGs to create distributed representations of the nodes, so that entities related to each other (e.g., having the same entity type) in the KGs are closer in the embedding space.
KGs can be derived from the database at hand, or they can be taken from external sources, similarly to distant-supervision approaches. For example, \new{the database in our deployment} contains information about companies and their relations (e.g., in terms of ownership and subsidiaries). 
Databases that store information about businesses explicitly or implicitly incorporate relations between companies and their attributes as, for example, described in works~\cite{mirylenka2016applicability, mirylenka2016recurrent, mirylenka2019hidden}. 
This means that companies and their connections can be represented as a graph and stored accordingly. 
As a simple example,  Figure ~\ref{fig:linkage-example} depicts graph modeling from a reference database, and shows how company entities, extracted from news articles, are linked to corresponding records.

Using graph representations for both structured and unstructured data sources makes it possible to leverage graph neural networks, thereby allowing all the available context to be used 
 in order to propagate useful information from one node to another. By context we mean all the attributes, connected records and sometimes external information that allows to fully understand and disambiguate entities.
 
The propagation of useful information from a node to its neighbors allows us to build discriminative node embeddings, which are necessary for higher-level tasks. Recently, graph neural networks (GNNs) yielded promising results for modeling hidden representations of graphs. They have already achieved state-of-the-art performance on several tasks, including link prediction, node classification, and graph generation~\cite{Guyon2017, Jiaxuan2018, Jiaxuan2018NIPS, Zhang2018, Ying2018, Xu2018, Velickovic2018, Xu2019}. 
Modeling entity representations through graphs and building their embeddings using GNNs has the potential to greatly improve performance in the field of data integration as well.

In general, KGs can model different kinds of relations (e.g., companies owning other companies, operating in the same domain, or having headquarters in the same city), and some relations can even be implicit (e.g., by considering transitive closure or by taking into account the distance between nodes in the graph).  
For simplicity, we focus on a single type of relation, but the same approach can be extended to consider multiple relations among entities.

Our approach is based on \new{combining Siamese and Graph Convolutional Networks (S-GCN)} to learn a distance function between the nodes.
At training time, the network is optimized to produce small distances for pairs of nodes belonging to the same entity, and large distances for unrelated nodes.
Therefore, the trained network can then be used to assess the similarity between a query and any reference node.
S-GCNs are especially useful when the task involves a large number of classes, but only a few examples for each label are available.

\new{One of the most important properties of graph modeling and S-GCNs is their ability to catch important information from the data and their structure. It means that the similarity and dissimilarity of objects can be fully defined by their initial properties and connections, so that only a small amount of labeled data is needed to build informative object representations. This property makes graph modeling with S-GCNs extremely suited for relational databases, as they provide important structural information by definition.}

The main contributions of this paper are as follows:\\
\begin{itemize*}
\item  we investigate how graph modeling and GNNs can be applied to data integration problems;\\
\item we propose a general approach for modeling hidden representations of entities in structured and unstructured databases;\\
\item \new{we devise various architectures of Siamese Graph Convolutional Networks for data-linkage tasks and evaluate the performance of the models for high scale record linkage tasks over business entity databases.} 
\end{itemize*}


\section{Related Work}
\label{sec:related_work}

Although the idea of developing and studying neural networks capable of manipulating and  processing graph-structured inputs dates back more than ten years \cite{Gori2005,Scarselli2009}, these kinds of models have recently been rediscovered and gained an unprecedented popularity for several different tasks and domains \cite{Battaglia2018}.
Such architectures have proven useful for many problems that require reasoning on a set of discrete entities, such as combinatorial optimization \cite{Khalil2017} and satisfiability \cite{Selsam2018} problems. 
More recently, \cite{Kool2019} successfully applied a model based on the Transformer architecture \cite{Vaswani2017} to solving the routing problems on graphs, such as the well-known traveling salesman problem (TSP).

Some of the most interesting and promising contributions in this field
include message-passing neural networks \cite{Gilmer2017} and non-local neural networks \cite{Wang2018}. In addition, the graph networks introduced by \cite{Battaglia2018} generalize several previous works, thereby providing a fundamental building block for applying deep learning to structured knowledge.

Recently, GNNs have been exploited for different facets of the entity linking tasks. For example, LinkNBed~\cite{Trivedi2018} is a deep relational learning framework for link prediction in multiple KGs. In that work, joint inference is performed over multiple KG structures that capture contextual information for entities and relations. To mitigate the limited amount of training data, they propose using multitask learning by leveraging additional kinds of data in different scenarios, such as unlabeled and negative instances. In contrast, our framework proposes a distance learning approach together with GCNs to overcome the lack of training data in a consistent way.

Another GNN framework, namely GraphUIL, has been introduced for user identity linkage in~\cite{Zhang2019}. GraphUIL ensures that both local and global information forming the user's social graph is propagated in order to jointly learn useful representations to identity linkage.  

Learning representations for the nodes in a graph proved beneficial for various tasks, such as node classification, clustering and link prediction~\cite{Xu2018, Monti2017}. 
GNNs learn node representations through iterative signal propagation between the hidden features of the neighboring nodes. 
As shown in~\cite{Xu2018},  $k$ iterations of the signal propagation process  allow reaching the entire subtree of depth $k$ rooted at a given node. This learning process generalizes the Weisfeiler--Lehman graph isomorphism test~\cite{weisfeiler1968}, indicating that both topology and hidden features of the neighborhood are learned simultaneously~\cite{Shervashidze2011}. 

State-of-the-art methods for entity matching of unstructured data have recently applied various neural network architectures to learn entity representations.
For instance, \cite{Ganea2017} describes how an attention mechanism over local context windows generates proper embeddings for document-level entity disambiguation.
A novel zero-shot entity matching task with previously unseen entities and a limited amount of training data for specialized domains is introduced in~\cite{Logeswaran2019}.
The goal of that work is to build a linker that can generalize to unseen specialized entities from unstructured sources with only textual context being available.
Best results have been obtained by a model based on BERT \cite{Devlin2019},  which was pre-trained on a specialized domain corpus and then fine-tuned using a small amount of the available training data.

Recently,~\cite{Ma2018} and \cite{Ktena2017} have combined GCNs and Siamese neural networks to perform the graph-matching task on functional brain networks.
These approaches extract graphs from corresponding MRI images and learn how similar two brain networks are. A recent work \cite{Liu2018} has proposed an approach for matching long text documents via training Siamese GCNs on their corresponding graph representations. The authors also presented a method for building a graph representation of a document, in which nodes are different concepts (keywords) and edges are interactions between concepts. One of the most closely related works \cite{Fakhraei2019}, though not on graphs, demonstrated the potential for learning embeddings for entity matching. Siamese networks have been employed to create semantic embeddings for protein and chemical names, and the linkage algorithm then simply performed a 1-NN search in the embedding space.

\new{Our work is different from the studies above for the following aspects:
\begin{enumerate*}[label=\itshape{(\roman*)}]
\item we focus on entity linkage rather than graph matching problems;
\item in our problems, query data can be represented by just one node (or several nodes if additional context information is available) rather than a complete network of a brain, or graph of concept interaction between sentences in a large document; 
\item we aim at working with thousands of distinct entities rather than at distinguishing disordered and healthy subjects or learning the similarity of a few 3D image classes.
\end{enumerate*}
}

The expressive power of GNNs is analyzed theoretically in~\cite{Xu2019}. The results of this study show that GCNs have limited discriminative power and cannot distinguish some changes in the graph structures. Though a limitation in the general case, this property is beneficial for record linkage tasks.
There is usually some discrepancy between the query entity and the reference database, 
as, due to the limited context availability, the query entity  lacks certain links compared to the more complete representation in the reference database.
Also, \cite{Kipf2017} experimentally shows that GCNs with two or three layers perform best for the node-classification task, outperforming deeper configurations.
This may be partially connected to the results obtained in the late 1980s, called the universal approximation theorem, which showed that a network with a single hidden layer can approximate any continuous function having compact support with arbitrary accuracy as the width of the layer tends to infinity~\cite{cybenko1989, Lin2018}. Thus, both shallow and wide networks are universal approximators. 
The approximation properties of deep and narrow networks were studied in~\cite{Lin2018}, which demonstrated that this kind of networks, having a single neuron per hidden layer, are sufficient to provide a universal approximation as the network depth tends to infinity.
\new{These approximation guarantees bring additional understanding of the capabilities of GCNs, and deep networks in general, for interpolating complex decision boundaries that appear in data integration tasks.}

\section{Problem Statement}
\label{sec:problem_statement}
\label{sec:problem}

\begin{small}
\begin{table*}[t]
\centering
\caption{Notation}
\label{tab:notation}
\begin{tabular}{|l|l|}
\hline
$E$ & set of entities \\ \hline
$R, Q$ & databases defined over entities in $E$ \\ \hline
$r \in R, q \in Q$ & a record in $R$ or $Q$, respectively \\ \hline
$e_r, e_q \in E$ & entities associated with records $r \in R$ and $q \in Q$, respectively \\ \hline
$R^*, Q^*$ & graphs associated with databases $R$ and $Q$, respectively \\ \hline
$r^* \in R^*, q^* \in Q^*$  & nodes associated with entities $e_r$ and $e_q$, respectively \\ \hline
$R^*_r, Q^*_q$ & sets of the neighboring nodes of $r^* \in R^*$ and $q^* \in Q^*$, respectively \\ \hline
$\Gamma_R, \Gamma_Q \subset \mathbb{R}^M$ & sets of embeddings generated from each node $r^* \in R^*$ and $q^* \in Q^*$, respectively \\ \hline
$\gamma_r \in \Gamma_R, \gamma_q \in \Gamma_Q$ & embeddings of nodes $r^*$ and $q^*$, respectively \\ \hline
$\Gamma_q^k \subseteq \Gamma_R$ & $k$ nearest neighbors of the embedding $\gamma_q$ in $\Gamma_R$ \\
\hline
\end{tabular}
\end{table*}
\end{small}

We assume to have two different databases $R$ and $Q$, defined over a set of entities $E$, such that each record $r \in R$ and $q \in Q$ can be regarded as an entity mention and uniquely associated to an entity $e \in E$.
Each entity represents a distinct real-world object, and we denote with $e_r$ and $e_q$ the entities associated to record $r \in R$ and record $q \in Q$ respectively.
The goal of record linkage is to find a function $\mathcal{F}$ that takes as input the databases $R$ and $Q$, and outputs a set of matches $\mathcal{M} \subset Q \times R$, such that $(q, r) \in \mathcal{M} \iff e_q = e_r$. Assuming we are matching records in $Q$ against records in $R$, we will henceforth refer to $R$ as the reference database.

In this work, we envision that research in the area of data integration could benefit from following a novel workflow designed to leverage the underlying relations among records and attributes in the database.
More specifically, we propose to extract graph representations of the databases and exploit the self-supervision provided by the graph structure to  train machine learning (ML) model for data integration, without the need for human-labeled data.
To this end, we divide the record linkage problem into two subtasks:
\begin{enumerate*}[label=\itshape{(\roman*)}]
\item graph modeling and 
\item GNN design.
\end{enumerate*} 

The graph modeling stage is about extracting two graphs $R^*$ and $Q^*$ from the databases $R$ and $Q$. 
Note that $Q$ is not necessarily produced from structured sources, as the data in $Q$ can come from free text, such as news articles.
Given any two records $r \in R$ and $q \in Q$,  we denote their corresponding graph nodes as $r^* \in R^*$ and $q^* \in Q^*$, respectively.

A node $r^* \in R^*$  represents a record $r \in R$  and corresponds to a real-world entity $e \in E$.
For instance, in Figure~\ref{fig:linkage-example}, $R$ is the reference database and $Q$ is a database extracted from news. Graph nodes  $r^*$ and $q^*$ are modeled via the attribute \textit{``Company''}, so the record $r_0 \in R$ with $ID=0$ is represented by the node $r^*_0 \in R^*$, that refers to the entity \emph{General Motors LLC}. Edges between the nodes are modeled via the attribute \textit{``Headquarter''}, so $r^*_1$ (\emph{General Motors China, Inc.}) has an edge to  $r^*_0$ (\emph{General Motors LLC}).
In certain cases, different nodes can refer to the same entity.
\new{The edges of the graphs are defined through the relations provided by the initial databases $R$ and $Q$. Different types of relations determine different strengths of the information flow from one node to another. The ``bandwidth'' of the edges can be defined a priori or learned automatically during the GNN training phase.}

%




The second stage is to design a neural network that can  be trained natively on the graph extracted from the reference database in order to learn a function $\mathcal{N}$, such that:
\[
\mathcal{N}(q^*, R^*) = 
\begin{cases}
r^* & \text{if } \exists r^* \in R^* : e_r = e_q \\
\bot & \text{otherwise} ,
\end{cases}
\]
where $\bot$ means that the node $q^*$ could not be matched to any node $r^* \in R^*$.
Note that $r^*$ is a single node, but in general  $\mathcal{N}(q^*, R^*)$ can produce several matching nodes from $R^*$ representing the same entity $e_r = e_q $. Following the example above, the node $q^*_1 \in Q^*$ (\emph{GM}) is matched to $r^*_0$ as they both represent the same entity \emph{General Motors LLC}.

Table~\ref{tab:notation} summarizes the notation introduced in this section and provides some additional notation that will be covered in the remainder of the paper.

\section{Proposed Approach}
\label{sec:approach}

We approach record linkage as a two-stage process. 
The first stage consists of obtaining a graph representation $R^*$ of the reference database $R$. The second stage involves designing and training a GNN on $R^*$   to perform entity-matching tasks on graphs. In the following, we describe our vision of the entity-matching process and propose the S-GCN (Siamese Graph Convolutional Network) model.

This paper assumes we can obtain a  graph representation of the entities of interest.
This representation can be extracted in many ways, either by exploiting structures in the database (links between entities, such as foreign keys) or by using an external source. 
Several methods have also been introduced in the literature for extracting knowledge graphs from both unstructured data \cite{Das2018, Gangemi2016} or semi-structured data \cite{Lehmann2015}.
The main contribution of this work lies in the way the graph structure is leveraged and processed to perform data integration.



\subsection{GNNs for Data Integration}
\label{sec:gnns_for_data_integration}

\begin{figure*}[ht]
 \begin{center}
   \includegraphics[width=0.85\textwidth]{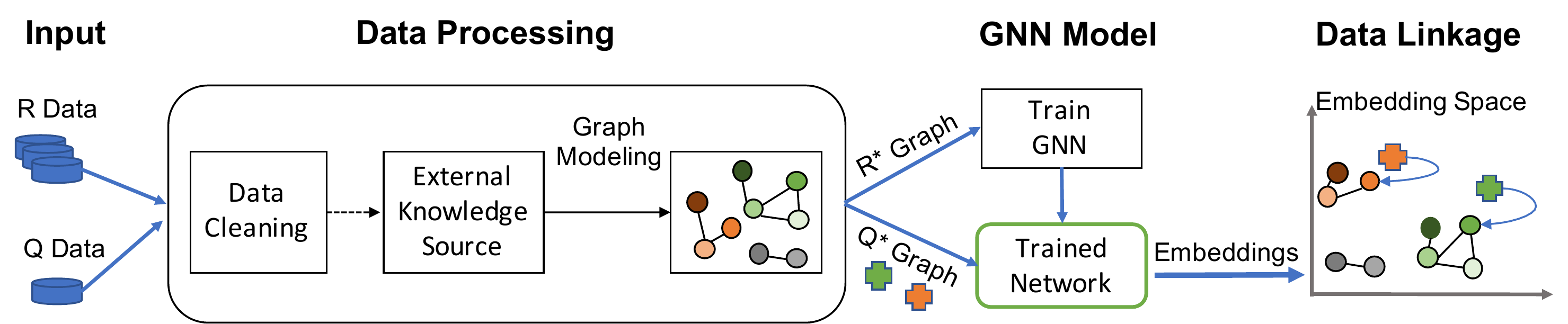}
  \caption{Entity matching workflow with graph neural networks}
  \label{fig:workflow}
  \end{center}
\end{figure*}

 

The reference graph $R^{*}$ is defined as a directed multigraph, and we denote with $|R^{*}|$ the number of nodes. We assume to have a feature descriptor $x_r \in \mathbb{R}^D$ for every node  $r^* \in R^*$, and we aggregate all these features in a matrix $X_R \in \mathbb{R}^{|R^{*}| \times  D}$.
If nodes include textual elements, such as names or descriptions, then features can be extracted using word embeddings, like Word2Vec \cite{Mikolov2013} and GloVe \cite{pennington2014}, or recent contextual language models like BERT~\cite{Devlin2019}.
Edges and relations for every pair of connected nodes are defined as triples of the form $(r^*_{i}, \rho ,r^*_{j})$, where $\rho$ is a relation type, while $r^*_i$ and $r^*_j$ are nodes of the graph. On the other hand, a query graph $Q^*$ consists of nodes  $q^* \in Q^*$  that have to be matched  to nodes in  $R^*$.

Our goal is to correctly match a node $q^* \in Q^*$ to a node $r^* \in R^*$ such that $e_q = e_r$. 
In order to achieve this goal, we propose to process the graph with GNNs, as they allow leveraging relations between connected nodes, by exploiting the self-supervision provided by the structure of the reference graph. This process can be implemented in several ways. 
The first approach is to adopt the conventional classification setup. 
In that case, the GNN model can be extended with a final softmax layer, where the number of output nodes is equal to the number of classes (the number of unique entities in the set of entities $E$). 
However, this kind of models usually requires a large number of training examples for each class label.
In contrast, for record linkage, the number of unique entities (classes) is typically much larger than in classification problems, and  the number of records belonging  to the same entity is usually small.
Also, this approach requires re-training the model when new entities are added to the database.


The methodology we propose is instead based on learning distributed representations for every node in $R^*$, so that nodes representing the same entity are closer in the embedding space and far from irrelevant nodes. This can be achieved by applying \textit{deep metric learning} methods, such as Siamese networks \cite{Chopra2005} or triplet networks \cite{Schroff2015}. $R^*$ is used as training data for a GNN that produces an embedding vector of size $M$ for each node $r^* \in R^*$. We denote as $\gamma_r$ the embedding of the node $r^* \in R^*$, whereas $\Gamma_R$ represents the set of all the embeddings for every node in $R^*$.
At inference time, for a node $q^* \in Q^*$ we compute an embedding vector $\gamma_q  \in \mathbb{R}^M$ by applying the already trained GNN model.
At this point, we have three potential scenarios:
\begin{enumerate*}[label=\itshape{(\roman*)}]
\item $q^*$ cannot be matched successfully to any node in $R^*$;
\item $q^*$ can be matched to a single node $r^* \in R^*$;
\item $q^*$ can be matched to multiple nodes that refer to the same entity.
\end{enumerate*}

For the first two cases, the rule to link $q^*$ to a node in the reference graph is given by:
\begin{equation}
	 r^*_c = \arg \underset{r^* \in R^*} \min dist(\gamma_{q}, \gamma_{r}), 
	\label{def:closest}
\end{equation}

\begin{equation}
	\mathcal{N}(q^*, R^*) = 
	\begin{cases}
	 r^*_c & \text{if } dist(\gamma_{q}, \gamma_{r^*_c} ) < t \\
	\bot & \text{otherwise,}
	\end{cases}
	\label{def:linkaGNN}
\end{equation}
where $dist(\gamma_{q}, \gamma_{r})$ is the distance (e.g., Euclidean) between vectors $\gamma_{q}$ and $ \gamma_{r}$, $ r^*_c \in R^*$ is the closest node to $q^*$ in the embedding space $\Gamma_R$, $t \in \mathbb{R}$ is a threshold on the distance, and $\bot$ means no matches were provided according to the given threshold $t$.

If multiple nodes can refer to the same entity, the predicted probability of a match to an entity $e$ is proportional to the sum of the similarities between the embeddings of the $k$ nearest neighbors $\Gamma_q^k \subseteq \Gamma_R$ and the query embedding $\gamma_q$:
\begin{equation}
	P_k(e_q=e \mid q^*, R^*) \propto \sum_{\gamma_r \in \Gamma_q^k} \delta(e, e_r) \cdot (1 - dist(\gamma_q, \gamma_r)),
	\label{def:knn}
\end{equation}
where $\delta(e, e_r)$ is $1$ if $e = e_r$ and $0$ otherwise.
This is similar to $K$-nearest neighbors classification, where a new sample is labeled according to the majority vote within labels of its nearest neighbors in the training set. The workflow described in this section is depicted in Figure \ref{fig:workflow}.

\subsection{Siamese GCN for Record Linkage}
\label{sec:siamese_GCN_for_record_linkage}

In this section, we introduce a model capable of performing entity matching on graphs by combining the advantages of graph convolutional networks~\cite{Kipf2017} and Siamese networks~\cite{Bromley1993}. The model, termed Siamese Graph Convolutional Network (S-GCN), has the primary objective of  learning discriminative hidden representations of nodes in a graph $R^*$, in such a way that they can then be used for further entity matching with previously unobserved data.

\begin{figure}[H]
	\begin{center}
		\includegraphics[width=0.7\linewidth]{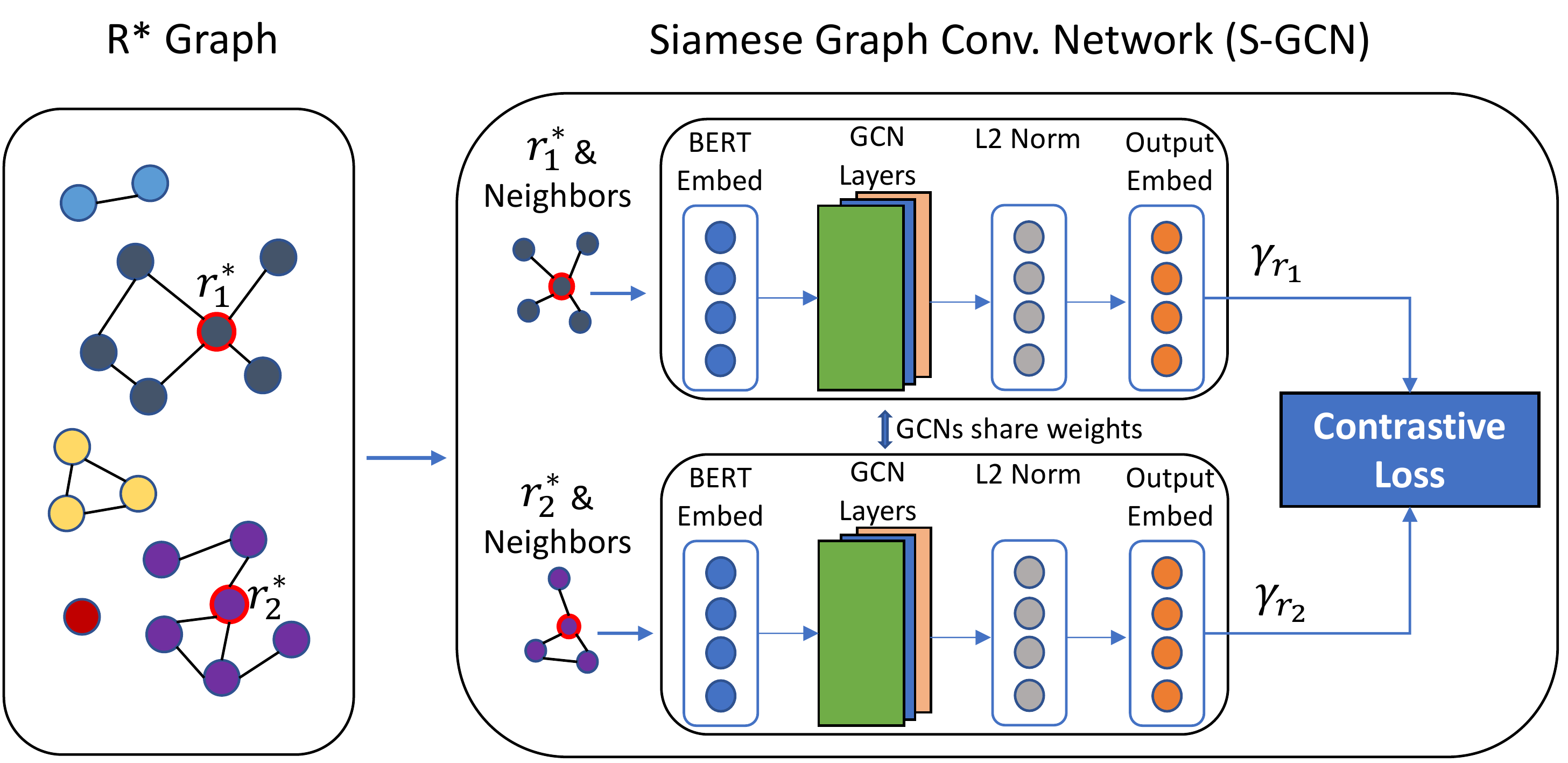}  
		\caption{Architecture of the siamese graph convolution network}
		\label{fig:sgcn}
	\end{center}
    \vspace*{-10pt}
\end{figure}

\textit{GCNs} \cite{Kipf2017} are a type of GNN that share filter parameters among all of the nodes, regardless of their location in the graph. Every GCN layer can be regarded as a non-linear function:

\begin{equation}
	Z^{(l+1)} = \phi(Z^{(l)}, A),
	\label{def:gcnLayer}
\end{equation}
where $l$ is the layer number and $A$ is the adjacency matrix of the graph $R^*$. $Z^{(0)}$ is given by the initial feature matrix $X_R$. In our experiments, we employed a pre-trained BERT model~\cite{Devlin2019} as the input layer to create input features for nodes. The last GCN layer $L$  produces the output embedding matrix for all nodes, which forms the set of embeddings $\Gamma_R$, where each embedding has dimension $z$. The propagation rule $\phi(\cdot)$ introduced in \cite{Kipf2017} is defined as:
\begin{equation}
	\phi(Z^{(l)}, A) = \sigma(\hat{D}^{-0.5} \hat{A} \hat{D}^{-0.5} Z^{(l)} W^{(l)}),
	\label{def:gcnProp}
\end{equation}
where $ \hat{A} = A + I$, $I$ is the identity matrix, $\hat{D}$ is the diagonal node degree matrix of $ \hat{A}$, and $W^{(l)}$ is a weight tensor of the $l$-th GCN layer.

\textit{Siamese neural networks} are a popular approach to implementing one (or few) shot learning algorithms, so that items from the same category are close in the embedding space and far from items of different categories.
A Siamese neural network is composed of two identical networks that share the same weights and accept two items as input. The first network outputs an encoding of the first item, and similarly the second one outputs an encoding of the second item.
Our Siamese network is based on two identical GCNs. During the training period, the first GCN focuses on a given node $r^*_1$ and its local neighborhood nodes $R^*_{r_1} \subset R^*$. This means that the GCN takes a subgraph around $r^*_1$ as input and produces a vector of size $z$ that finally forms the embedding vector $\gamma_{r_1}$ of node $r^*_1$. The same procedure is repeated with the second GCN for node $r^*_2$, as shown in Figure~\ref{fig:sgcn}. Siamese networks are optimized with respect to the loss between their two outputs, namely $\gamma_{r_1}$ and $\gamma_{r_2}$. The loss will be small for two nodes representing the same class (e.g., nodes of the same entity) and will be greater for nodes belonging to different classes. An instance of such a loss is the contrastive loss~\cite{Hadsell2006}:

\begin{equation}
\begin{aligned}
&Contrastive Loss (\gamma_{r_1}, \gamma_{r_2}) = \\
=& 0.5 \cdot (1+y) \cdot dist(\gamma_{r_1}, \gamma_{r_2})^2 + \\ 
+& 0.5 \cdot  (1-y) \cdot \{max(0,m-dist(\gamma_{r_1}, \gamma_{r_2}))\}^2,
\label{def:contrLoss}
\end{aligned}
\end{equation}
where $y = 1$ if nodes $(r^*_1, r^*_2)$ are from same class, and $y=-1$ otherwise, while $m$ is a margin that is a non-negative number and indicates that dissimilar node pairs whose distance is beyond the margin will not penalize the network.

When the model is trained,  one of the two Siamese networks (identical GCNs) is used for computing the output embedding matrix and forms $\Gamma_R$ according to \eqref{def:gcnLayer}. It then sends the whole graph $R^*$ as input to the GCN. Similarly, using the trained GCN, we can compute $\Gamma_Q$ for an unobserved graph $Q^*$ and perform entity matching in accordance to \eqref{def:linkaGNN} and~\eqref{def:knn}.

\subsection{Siamese GCN with Subspace Learning}
\label{sec:siamese_GCN_with_subspace_learning}
\new{Training S-GCN with a large number of entities can suffer from poor local minima, as selecting informative dissimilar pairs of entities becomes extremely hard and computationally expensive. Inspired by~\cite{Wang2017H}, we adopt the subspace learning (SL) approach to train  our Siamese GCN more effectively. In the following, we will refer to this network as S-GCN$_{SL}$.}

\new{A generally good way to train neural networks for distance metric learning is to select hard instances of dissimilar entities. In our case, the most informative pairs of dissimilar entities are given by nodes whose representations are close in the hidden space, but they refer to different entities. Therefore, sampling such difficult and informative entities leads the network to learn how to handle ambiguity and generalize better. However, as the number of entities grows, it becomes computationally intractable to find the most informative dissimilar pairs at every learning iteration. The idea of training S-GCN with SL is to initially generate the representation of all entities (e. g., via S-GCN) and then cluster all entity representations to form subspaces of the most similar entities. Then, the model is trained again over all subspaces, and dissimilar pairs are selected randomly within a subspace only.  After each epoch, the updated model is used to generate new representations of the entities and the clustering and training steps are repeated again.}

\newcommand{\duns}{D-U-N-S}

\section{Datasets and Task Setup}
\label{sec:datasets_and_task_setup}

To investigate the feasibility of our proposed approach, we set up a data integration task on a proprietary relational database that contains fine-grained information about business entities. 
This section describes how these data have been modeled as a graph and how the record-linkage task has been prepared and set up.

\subsection{Graph Extraction}
\label{sec:graph_extraction}

Each record in the 
 database corresponds to a different company business location and is assigned a  unique local 
identifier
that can be used to track  business entities precisely.
Company entities represented in the database include different branches, subsidiaries and headquarters, each of which reports directly or indirectly to a given global identifier. 
Hence, companies related to the same global identifier can be grouped together and represented hierarchically as a single family.
In addition, for each company entry, 
we have a textual name, and, optionally, up to three alternate names (aliases).
 
As mentioned in Section~\ref{sec:problem}, we are interested in linking a company name to a family in the 
 database. To this end, we represent families in the 
  database as a graph, where each node is a company name.
Relations between nodes are derived from the hierarchical representation of 
company families and from the aliases provided in the database.
More precisely, a relation between two nodes exists if they are adjacent business entities in the  
hierarchy defined by the database, or if they are aliases of each other.
Hence, the resulting graph consists of several disjoint connected components, each corresponding to a different 
 family.
The 
company database also provides other potentially valuable information, such as the precise location of each business entity.
In this task, however, we do not model this additional information because we want to link companies based only on their names.
This becomes useful in several real-world scenarios, as the company name that needs to be matched to the database may come from different structured or unstructured data sources, where this kind of additional information is usually not be available.

\subsection{Data Preprocessing and Partitioning}
\label{sec:partitioning}

Overall, the 
company database contains over 100M 
business entities located in approximately 200 countries.
We focus our experiments on the 
records corresponding to companies located \new{either in Switzerland or in the United States.
This accounts for around 700k 
and 27M 
entities, respectively. We tested our approach on different graphs extracted from both datasets.} 

As discussed in Section~\ref{sec:graph_extraction}, the graphs extracted from the company 
database contains a disjoint subgraph for each family.
To investigate the feasibility of our approach, we randomly remove nodes from the graph, and we then use the trained S-GCN model to predict the subgraph a previously unseen company name belongs to.
More specifically, we partition our original dataset into training, validation, and test sets.
The training set contains company names and specifies aliases and relations between entities in the same 
family.
The validation and test sets  include only the name of a company that is not represented in the training set and a class label that refers to the correct 
 family. 
This means that the training set contains a graph for each 
company family, whereas the test set contains only \textit{isolated nodes}.
Modeling this kind of scenario with GNNs is challenging because nodes in the test set lack contextual information that would  usually be derived and aggregated from neighboring nodes. 
Section~\ref{sec:short_names} describes a possible way to mitigate this problem in our use case.

To provide enough training examples for each class label (i.e., for each 
company family), before partitioning the data into training, validation, and test sets, \new{we first remove company families with less than $t \in \mathbb{N}$ companies. The threshold $t$ is set to $5$ for the graph that represents companies located in Switzerland, whereas it is set to $10$ for the United States.
This preprocessing step allows increasing the variability of the company names processed by the model at training time and guarantees that each family is adequately represented in the training, validation, and test sets.}

As the record-linkage task we have set up works at the family level, we do not have an exact match at the record level from an entity in the test set to an equivalent entity in the training set.
To address both issues, we partition the dataset into training, validation and test sets in such a way that names in the training set have \new{at least some} overlap with names in the test and validation sets.


\begin{table}[!htb]
\caption{Statistics about the datasets}
\label{tab:data}
\begin{center}
\begin{tabular}{|l|l|l|l|}
\hline
 & \textbf{overlap-1} & \textbf{overlap-all} & \textbf{\new{US data}} \\ \hline
\# of families & 900 & 700 & 28.6k \\ \hline
\# of nodes & 9k & 4k &  1.7M \\ \hline
\# of edges & 11k & 4k & 2.6M \\ \hline
avg. node degree & 2.5 & 2 & 3.2 \\ \hline
med. node degree & 1 & 1 &  2 \\ \hline
\begin{tabular}[c]{@{}l@{}}avg. \# of samples\\ per company family\end{tabular} & 10 & 6 & 116 \\ \hline
\begin{tabular}[c]{@{}l@{}}med. \# of samples\\  per company family\end{tabular} & 5 & 4 & 32 \\ \hline
\# of  val. nodes & 2k & 1.1k & 145k \\ \hline
\# of test nodes & 2k & 1.3k & 153k \\ \hline
exact matches in test set & 0\% & 0\% & 38\% \\ \hline
\end{tabular}%
\end{center}
\end{table}

More precisely, \new{for companies located in Switzerland,} we produce two different variants of the datasets, which we denote as overlap-1 and overlap-all.
In the overlap-1 variant, we impose the constraint that each name in the validation and test sets has at least one representative word in common with at least one training example that belongs to the same family.
This version of the datasets addresses the first problem of not having an exact match for company entities at the record level, while also mitigating the issue of non-overlapping names within the same family in the training set.
Intuitively, this variant is meant to ensure that we can always find a record-level match between an entity in the test set and an entity from the same family in the training set. 

On the other hand, in the overlap-all version, before partitioning the dataset into training, validation, and test sets, we first tokenize each company name into a list of words. We then identify a representative token for each family as the most frequent token among the names in that family.
Next, we filter out all names that do not contain the representative word for their family, and we partition the remaining names into non-overlapping training, validation, and test sets.
This process addresses the issue of non-overlapping names within the same family, and produces cleaner datasets that are more suitable for training our S-GCN model.
\new{The overlap-1 and overlap-all variants remain still challenging for both traditional approaches and the proposed methodology, as will be demonstrated in Section~\ref{sec:evaluation}. However, the datasets are specifically designed to provide enough commonalities to allow machine-learning approaches to generalize effectively to unseen records. }

\new{In order to generate the datasets that correspond to companies located in the United States (US data), we apply the same assumption used for producing the aforementioned overlap-1 variant.
However, in this case, before partitioning the data into training, validation and test sets, we do not remove duplicate company names within the same family.
This means that both the test and validation sets for the US data contain some names that appear in the training set.
Overall, this design choice makes the datasets larger and more indicative of the performance of the model in a real-world scenario.
Indeed, in practice, whether $Q$ is a pre-existing relational database or a set of business entities extracted from news, the chances that a record $q \in Q$ matches exactly with another record $r \in R$ may not be negligible. In our use case, the procedure described above yields a test set where the percentage of exact matches against the training set is equal to $37.7\%$ of the records. With a total of approximately 1.7 million nodes, the graph extracted from the US data is particularly suitable to evaluate the scalability of our approach.
In all the datasets introduced in this section, the training set includes $60\%$ of the records in the database, while the remaining $40\%$ is divided equally between the validation and test sets.
}
Table~\ref{tab:data} presents selected statistics about the datasets introduced in this section.


\subsection{Enriching the Graph with Short Names}
\label{sec:short_names}

As mentioned in Section~\ref{sec:partitioning}, although the training set contains a connected component for each 
company family, the models are evaluated on a test set that contains company names as isolated nodes.
To address this issue, we generate an additional alias for each company name in the test set, and link this alias to the real name.
The enriched test set thus consists of small connected components of size two.
This enhancement helps the GNN to better leverage the graph structure and the local neighborhood of nodes.

We generate additional names that are short or normalized versions of a company name. Short names are extracted using conditional random fields (CRFs) as described in \cite{gschwind2019}.  CRFs are trained such that the most discriminative word or phrase in a company name is extracted as a short name. The importance of short names for linking company entities is also demonstrated in~\cite{Loster2017}.


The same process can be applied to names in the training set in order to increase the size of the graph and provide more variability within the training data.
Table~\ref{tab:short_names} lists the number of distinct additional names that are added to the training, validation, and test sets by enriching the graphs with short names.

\begin{table}[!htb]
\caption{Number of distinct additional names added to the graph}
\begin{center}
\label{tab:short_names}
\begin{tabular}{|l|l|l|l|}
\hline
& \textbf{overlap-1} & \textbf{overlap-all} & \textbf{US data} \\ \hline
\textbf{Training set} & 5.2k & 1.9k & \new{240k}\\ \hline
\textbf{Validation set} & 1.7k & 0.9k & \new{90k} \\ \hline
\textbf{Test set} & 1.8k & 0.9k & \new{90k} \\ \hline
\end{tabular}%
\end{center}
\end{table}

 
\section{Experimental Evaluation}
\label{sec:evaluation}
This section provides an evaluation of the approach proposed in Section \ref{sec:approach}.
\new{We compare our approach against strong baselines, including a recent rule-based system \cite{gschwind2019} explicitly designed for the data described in Section~\ref{sec:datasets_and_task_setup}.
Then, we evaluate the quality of the embeddings learned by the S-GCN model.}


\subsection{Baselines}
\label{sec:baselines}

The approach outlined in Section~\ref{sec:approach} has been compared against two baselines.
The first baseline is a record linkage system (RLS) for company entities, recently introduced in~\cite{gschwind2019}.
This system relies on a rule-based approach to score entities with different attributes, including the name of a company, the precise location, Standard Industry Codes (SIC)\footnote{\url{https://www.osha.gov/pls/imis/sicsearch.html}}, and even short names, extracted as previously mentioned in Section~\ref{sec:short_names}.

We re-implemented the approach using the scoring function described in~\cite{gschwind2019}, which is based on a combination of weighted Levenshtein and Jaccard similarities between company names.
RLS provides matches at the record level. Given a record $q \in Q$, the system outputs a record $r_q \in R$ such that
\begin{equation}
\label{eq:rls}
r_q = \arg \max_{r \in R}(score(r, q)).
\end{equation}
where $score$ is the aforementioned scoring function detailed in \cite{gschwind2019}.
To adapt this method to our use case and perform matches at the family level, we simply define
$RLS(q, R) = f_{r_q}$, where $f_{r_q}$ is the family of the company represented by the record $r_q$, defined as in \eqref{eq:rls}.

The second baseline is a multilayer perceptron (MLP) with a softmax classification layer on top, optimized with a cross-entropy loss function. The MLP takes BERT features as input for company names and produces a company family as output. Hyperparameters have been optimized manually on the validation set. To evaluate MLP on the datasets, we used the following setup: 
\begin{compactitem}
	\item the number of fully connected layers varies from 1 to 3, depending on the dataset used;
	\item the learning rate is set to $10^{-3}$;
	\item every layer contains 500 hidden neurons;
	\item the dropout rate is 0.2;
	\item we apply a weight decay of $5 \cdot 10^{-5}$;
	\item the Adam optimizer is used for updating the weights of the model.
\end{compactitem}
We train the network for up to 1000 epochs and stop the training process if the validation loss does not improve in the course of 50 epochs.

\subsection{Record Linkage Evaluation}
\label{sec:rl-eval}

We use the accuracy score, i.e.~the proportion of correctly matched records in $Q$, to assess the performance of the proposed algorithms. As GCN is a well-established choice for classifying graph nodes, we designed a GCN with a final classification softmax layer as described in Section~\ref{sec:approach}.  Note that, since GCN-CLF is trained directly to perform classification, it must be retrained whenever a new company family enters the reference database $R$, whereas S-GCN is tolerant to new company families. S-GCN embeddings of nodes that represent a new company family will tend to group close to each other in the embedding space. This could allow the model to be extended to previously unseen families without having to train the network again from scratch.

\parg{Node features} 
Since nodes in our graphs correspond to company names, we can leverage well-established distributed representations of text in the form of word embeddings and contextual language models as meaningful initial features.
As company names may be written in different languages, consist of abbreviations and even non-existing words, 
we used character-level $n$-grams and the multilingual BERT model. First we applied WordPiece tokenization~\cite{Devlin2019} to company names and then fed the resulting set of tokens through the trained BERT model with the reduced mean pooling strategy to obtain features for company names.

\parg{Record Linkage with S-GCN} 
S-GCN is trained on the reference graph $R^*$ in a self-supervised way, so that the structure of the graph guides the training process. After the models have converged and all the reference nodes have been mapped into the S-GCN embedding space $ \Gamma_R $, we compute the embeddings  $ \Gamma_Q$ for all the nodes in $Q^*$ by feeding them through the trained S-GCN. Then a one-nearest-neighbor (1-NN) search is applied to find the most similar vector $\gamma_r \in \Gamma_R$ to a node $q^* \in Q^*$:
\begin{equation}
	\new{r^* =  SGCN(q^*, R^*) = \arg \min_{r^*_i \in R^*} dist(\gamma_{q}, \gamma_{r_i} )}.
	\label{def:1nn}
\end{equation}

As every vector $\gamma_r$ refers to an entity $e_r$ that in turn belongs to a family $f_r$, we can define the set of matches of records in $Q$ against records in $R$ as follows:
\begin{equation}
	\mathcal{M}_{SGCN} = \{ (q,r) \in Q \times R \mid SGCN(q^*, R^*) = \new{r^*} \}.
	\label{def:match}	
\end{equation}

\noindent
\new{Then, given a set of matches $\mathcal{M}_{\mathcal{N}}$ produced by a model $\mathcal{N}$, we define the accuracy as:}
\begin{equation}
	\new{Accuracy(\mathcal{M}_{\mathcal{N}}) = \frac{ | \{ (q,r) \in \mathcal{M}_{\mathcal{N}} \mid f_q=f_r  \} | }{| \mathcal{M}_{\mathcal{N}} |}}.
	\label{def:acc}
\end{equation}

\parg{Training} 
For every epoch, we generate training pairs of similar and dissimilar nodes. Two nodes are considered to be similar only if they are connected by an edge. Thus, some nodes cannot be joined into a pair of similar nodes, even if they belong to the same company family. \new{This helps the network to cope with families where some nodes may have completely non-overlapping names. As an example, in the overlap-1 dataset, the nodes \emph{Hotel Bellevue Palace} and \emph{Infracore SA} belong to the same family.} For every similar pair, we create 
dissimilar pairs by randomly sampling nodes from other company families.

\new{Hyperparameters have been optimized on the validation sets using grid search. For GCN-CLF on overlap-1 and overlap-all, we used the following setup: the number of GCN layers is 3 and 1 respectively; the learning rate is \new{$10^{-3}$}; every layer comprises 600 hidden neurons; the dropout probability is 0.2 and the weight decay is set to \new{$5 \cdot 10^{-7}$} and \new{$5 \cdot 10^{-5}$} respectively.} 
To evaluate S-GCN, we used the following setup: the number of GCN layers is set $1$; the learning rate is \new{$10^{-3}$}; there are $600$ hidden neurons in every layer; the dimensionality of the output embeddings is set to $300$; the dropout value is $0.2$; the weight decay is \new{$5 \cdot 10^{-6}$}; the margin value in the contrastive loss definition \eqref{def:contrLoss} is $0.45$ and $0.35$ respectively for overlap-1 and overlap-all; for every pair of similar nodes, two pairs of dissimilar nodes were sampled.
We then use the Adam optimizer to update the weights of GCN-CLF and S-GCN. We train the networks for up to $1000$ epochs and stop the training if the validation accuracy does not improve in the course of $50$ epochs.

\parg{Results}
\begin{table}[!t]
\caption{Accuracy of data linkage algorithms on the overlap-1 and overlap-all datasets}
\label{tab:res}
\begin{center}
\begin{tabular}{|l|l|l|}
\hline
 \textbf{Algorithm}& \textbf{overlap-1} & \textbf{overlap-all} \\ \hline
\textbf{RLS} & 0.71 & 0.80 \\ \hline
\textbf{MLP} & 0.71 & 0.90 \\ \hline
\textbf{GCN-CLF} & 0.56 & 0.90 \\ \hline
\textbf{S-GCN} & 0.70 & 0.91 \\ \hline 
\textbf{S-GCN$_{SL}$} & 0.75 & 0.92 \\ \hline  \hline 
\begin{tabular}[c]{@{}l@{}} \textbf{RLS} \\  (+short names) \end{tabular} & 0.78 & 0.87 \\ \hline
\begin{tabular}[c]{@{}l@{}} \textbf{MLP} \\  (+short names) \end{tabular} & 0.75 & 0.91 \\ \hline
\begin{tabular}[c]{@{}l@{}} \textbf{GCN-CLF} \\  (+short names) \end{tabular} & 0.70 & 0.94 \\ \hline
\begin{tabular}[c]{@{}l@{}} \textbf{S-GCN} \\  (+short names) \end{tabular} & 0.83 & 0.93 \\ \hline
\begin{tabular}[c]{@{}l@{}} \textbf{S-GCN$_{SL}$} \\  (+short names) \end{tabular} & \textbf{0.85} & \textbf{0.95} \\ \hline
\end{tabular}%
\end{center}
\end{table}
We compared the GCN-CLF and S-GCN models against the baselines on the datasets introduced in Section \ref{sec:partitioning}. In addition, we studied the effect of enriching the graphs $R^*$ and $Q^*$ with short names, as described in Section~\ref{sec:short_names}.
Overall, in all the datasets, the proposed graph-based approaches outperform the baselines. Analyzing the results on the overlap-1 dataset, we can easily see that enriching the graphs with short names allows improving accuracy by a large margin. Most notably, for S-GCN on overlap-1, accuracy rises by 13\%. 
Adding short names increases the variability of training data and allows GNNs to effectively propagate information from short versions to the original names. 

\new{It is worth mentioning that record linkage on overlap-1 is an inherently difficult task for machine-learning approaches, because the dataset contains ``noisy'' families, as mentioned in Section \ref{sec:partitioning}.
In this dataset, MLP performs better than GCN-CLF, but the best results are achieved by the S-GCN and S-GCN$_{SL}$ models, which reach an accuracy of 0.83 and 0.85 respectively. An explanation of the poor performance of GCN-CLF is given by the presence of connected non-overlapping company names in the KG. The network propagates this ``noisy'' information through edges thereby hampering the impact of ``clean'' samples.}

\begin{figure}[!b]
\begin{center}
\includegraphics[width=0.5\linewidth]{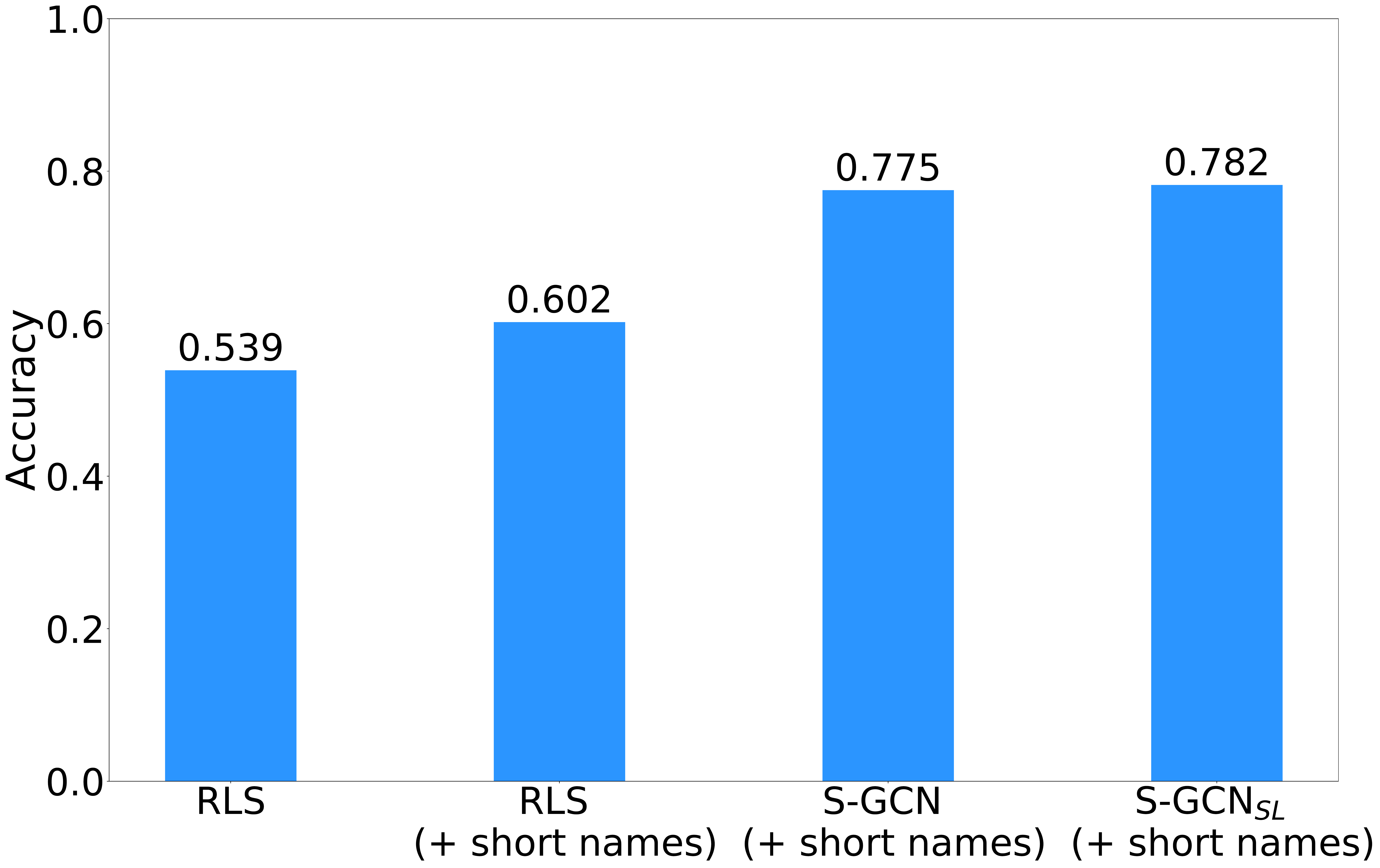}  
\caption{Accuracy on the US data}
\label{fig:us_results}
\end{center}
\end{figure}

\begin{figure*}[t]
\centering
\begin{subfigure}[t]{0.25\textwidth}
	\centering
	\includegraphics[width=.97\linewidth]{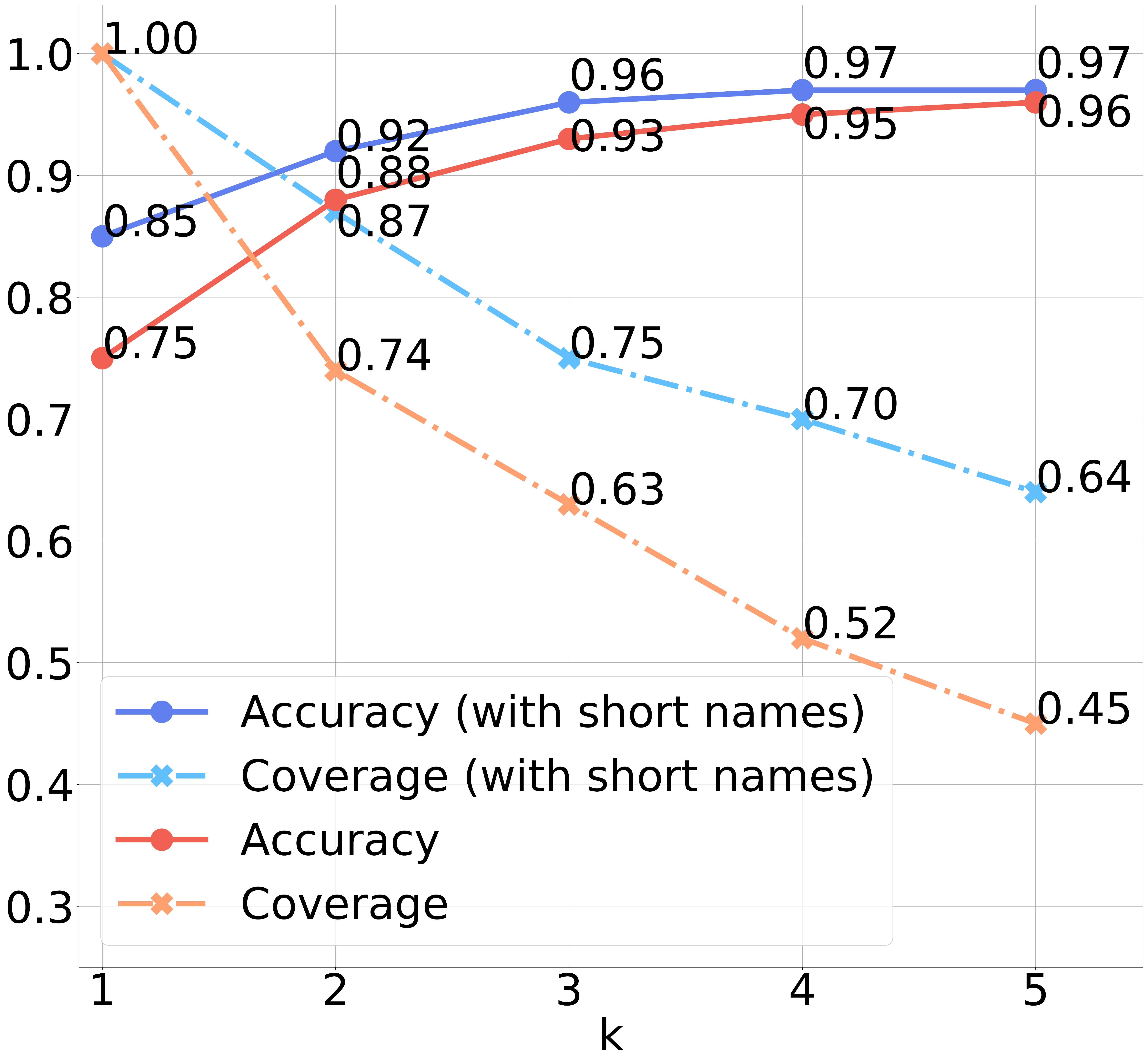}
	\caption{\new{S-GCN$_{SL}$ on overlap-1}}
	\label{fig:sgcn_overlap_1}
\end{subfigure}%
\begin{subfigure}[t]{0.25\textwidth}
	\centering
	\includegraphics[width=.97\linewidth]{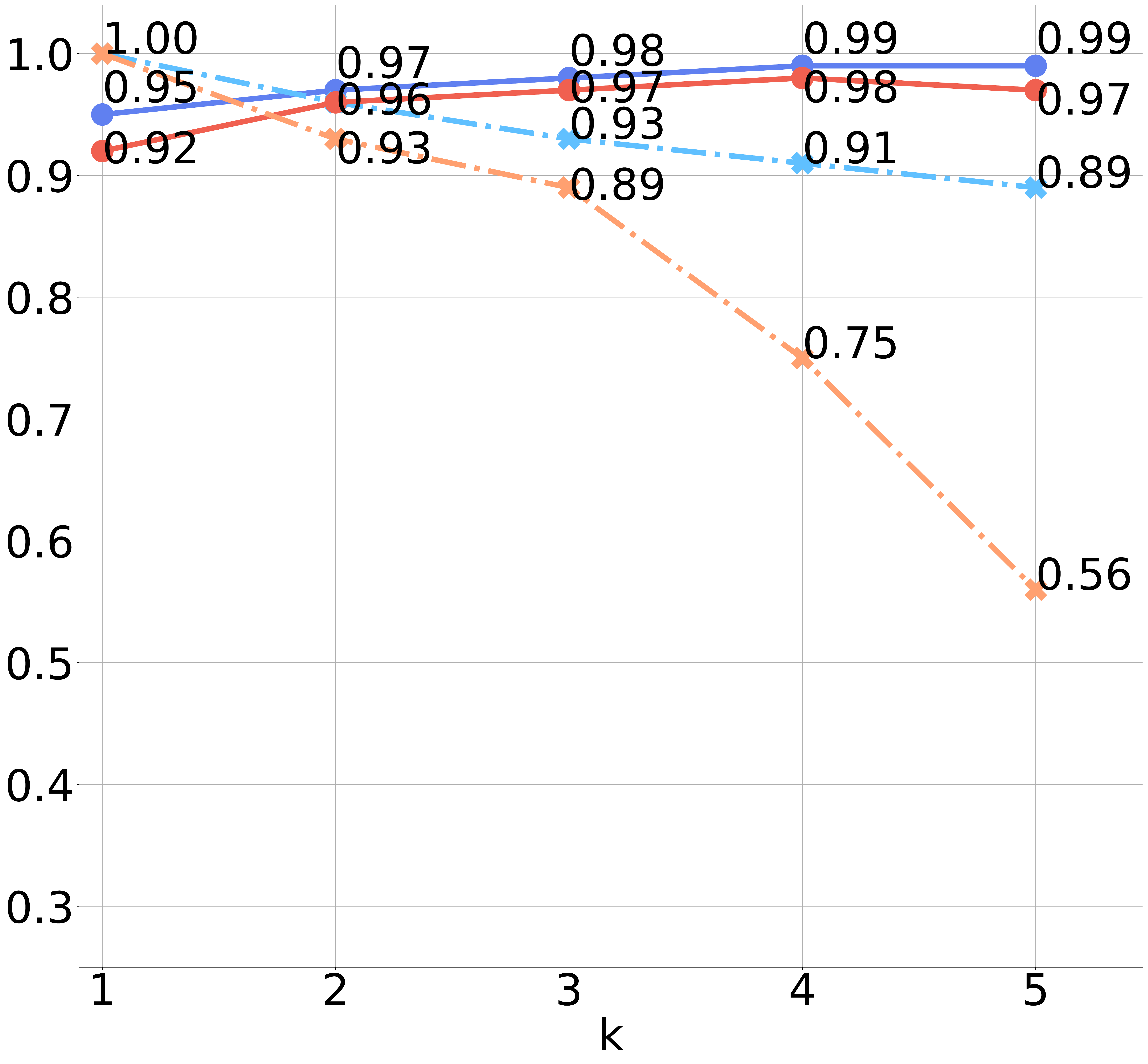}
	\caption{\new{S-GCN$_{SL}$ on overlap-all}}
	\label{fig:sgcn_overlap_all}
\end{subfigure}%
\begin{subfigure}[t]{0.25\textwidth}
	\centering
	\includegraphics[width=.97\linewidth]{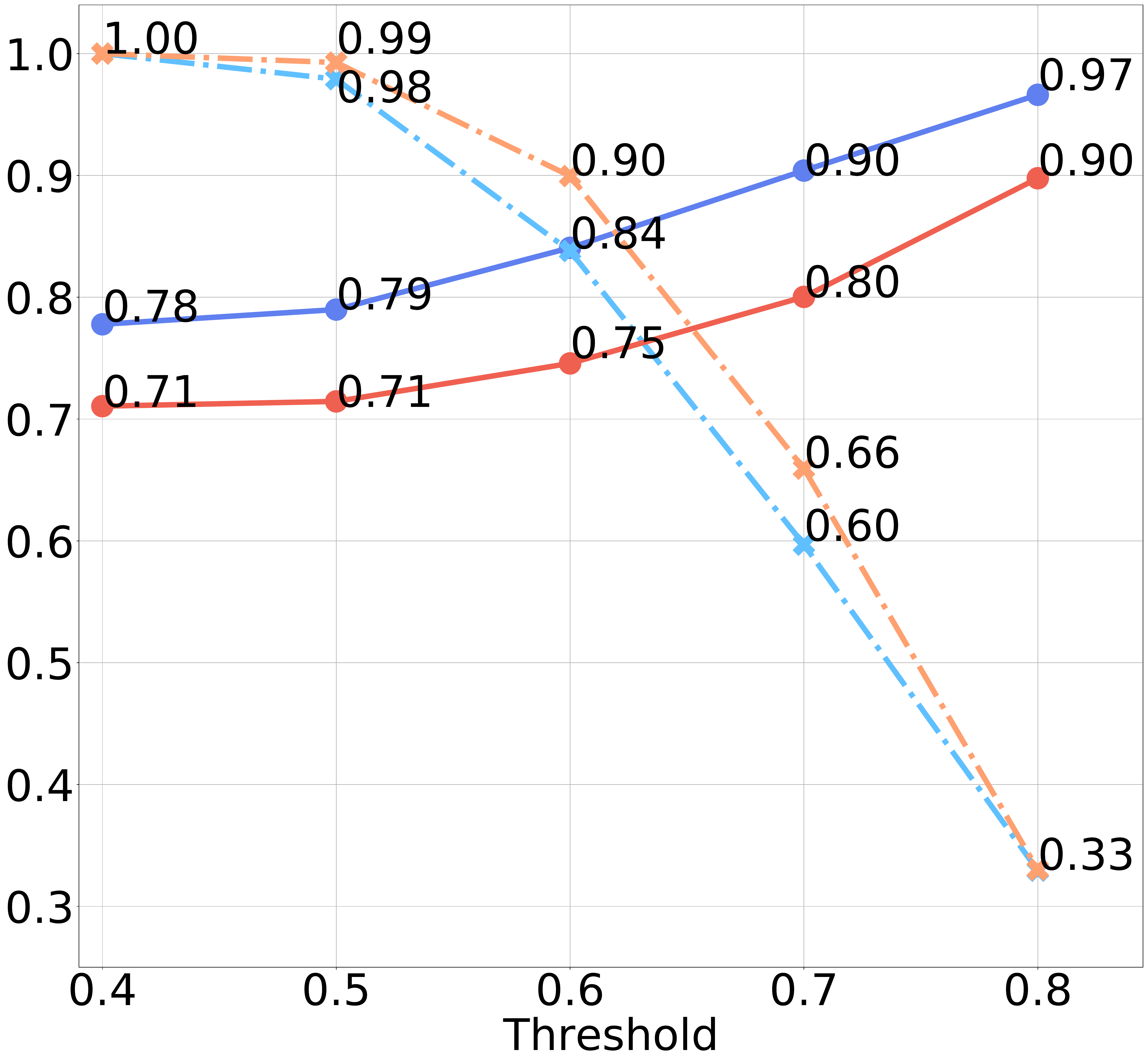}
	\caption{RLS on overlap-1}
	\label{fig:rls_overlap_1}
\end{subfigure}%
\begin{subfigure}[t]{0.25\textwidth}
	\centering
	\includegraphics[width=.97\linewidth]{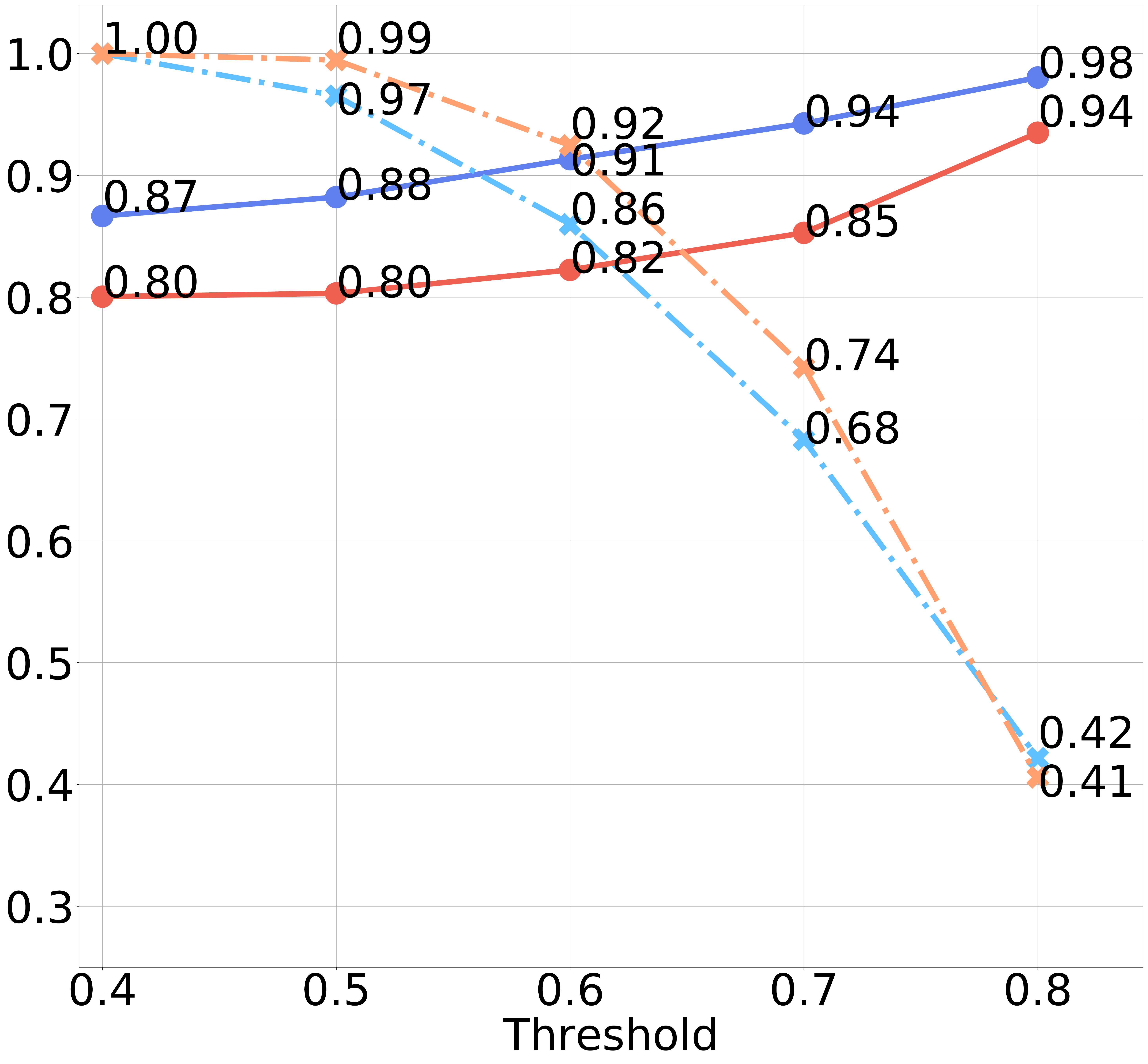}
		\caption{RLS on overlap-all}
	\label{fig:rls_overlap_all}
\end{subfigure}%
\caption{Accuracy and coverage obtained by \new{S-GCN$_{SL}$} and RLS on both overlap-1 and overlap-all}
\label{fig:acc_cov}
\end{figure*}

\new{S-GCN and S-GCN$_{SL}$ are less sensitive to noise within a company family and they outperform GCN-CLF by 13\% and 15\% on overlap-1 with short names. Recall that S-GCN is not trained directly for classification. Instead, it aims to put two similar nodes from $R^*$  closer within the embedding space $\Gamma_R$. We consider two nodes to be similar only if they are connected by an edge. Therefore, S-GCN tries to map to close embeddings only connected nodes, without explicitly taking into account the other entities in the same family. Although some noise might come from neighboring nodes, S-GCN does not aim to learn a common pattern among all these entities at once, and, subsequently, it does not explicitly force grouping all entities in the same family.}

\new{The overlap-all dataset is ``cleaner'', as it only contains company names that share a given representative word with all the other entities in the same family. Hence all algorithms achieve a higher accuracy on this variant. The highest performance is reached by approaches based on the use of GNNs. 
Accuracy levels for S-GCN, S-GCN$_{SL}$ and GCN-CLF are comparable and consistently exceed the results obtained by the baselines.}

\new{
Finally, we assessed the scalability of our approach by training our models a larger dataset that represents companies located in the United States.
As mentioned in Section \ref{sec:partitioning}, the training set contains 1.7 million nodes, grouped in more than 28k families.
On the one hand, the large size of the reference database is a potential threat for rule-based systems like RLS, but, on the other hand, dealing with a high number of classes poses hard challenges for approaches based on machine learning as well.
We trained on this dataset our top-scoring systems, namely S-GCN and S-GCN$_{SL}$, both with short names. Figure \ref{fig:us_results} summarizes the results. As we can see, without short names RLS does not achieve a satisfactory accuracy and is able to identify the correct match only for 54\% of the names in $Q$. The use of short names allows improving the performance of the system and achieving an accuracy of approximately 0.60. On this dataset, the S-GCN and S-GCN$_{SL}$ models achieve a comparable accuracy of 0.78, accounting for a significant improvement of $18\%$ with respect to the best baseline, namely RLS with short names. 
}



\subsection{Accuracy-Coverage Tradeoff}
\label{sec:tradeoff}


Both the S-GCN and RLS approaches allow the parameters of the system to be tuned such that a company name in the test set is linked to a family in the training set only if the match could be identified with high confidence.
In other words, for both systems we can specify a criterion to detect that a given company name in the test set could not be matched successfully to any family in the training set.
This increases the probability that the results provided by the system are correct, at the expense of some company names possibly being left unclassified, even if  a corresponding record exists in the reference database.
Hence, we can identify a tradeoff between the accuracy and coverage of the system, where  accuracy is defined by  \eqref{def:acc}, and coverage is simply the number of either correct or wrong matches provided by the system over the total number of queries (i.e., the size of the test set):
\[Coverage(\mathcal{M}, Q) = \frac{|\mathcal{M}|}{|Q|}.\]

RLS~\cite{gschwind2019} can be naturally tuned to avoid reporting wrong matches by simply setting a threshold $t \in [0, 1]$ on the value of the scoring function:
\[
RLS_t(q, R) =
\begin{cases}
f_{r_q} & \text{if } \text{score}(r_q, q) \geq t \\
\bot & \text{otherwise}, \\
\end{cases}
\]
where $score$ is the RLS scoring function and $r_q$ is an RLS match for $q$ against the database $R$, defined as in ~\eqref{eq:rls}.

Similarly, we can tune our approach based on S-GCN to link a node $q^* \in Q^*$ to a family $f \in F$ only if all the $k \in \mathbb{N}$ nearest neighbors of the predicted embedding $\gamma_q$ of $q^*$ belong to family $f$.
Formally,
\[
\textit{SGCN}_k(q^*, R^*) =
\begin{cases}
f & \text{if }  e_r \in f, \forall \gamma_{r} \in \Gamma_{q}^k \\
\bot & \text{otherwise} \\
\end{cases}
\]
where $\Gamma_{q}^k \subseteq \Gamma_{R}$ is the set of the $k$ nearest neighbors of the embedding predicted by our S-GCN model for $q^*$.

Figure~\ref{fig:acc_cov} shows the accuracy and coverage achieved on overlap-1 and overlap-all by both RLS and \new{S-GCN$_{SL}$} for different values of $t$ and $k$, respectively.
As outlined also in Table \ref{tab:res}, on overlap-1, at a coverage value of 1.00, without using short names, the accuracy of \new{S-GCN$_{SL}$ is only moderately higher than that of RLS}.
Clearly, increasing the values of $k$ and $t$ allows improving the accuracy of both systems at the expense of decreasing coverage.
However, for S-GCN we can improve accuracy while keeping coverage relatively high compared to RLS.
\new{Setting $k = 3$ is sufficient to exceed an accuracy of $90\%$ for S-GCN$_{SL}$, on overlap-1 without short names}. 
To achieve a comparable accuracy of $0.90$ for RLS, we have to increase the threshold to $0.8$, resulting in a very low coverage of $0.33$.

Using short names allows us to  boost the performance of \new{S-GCN$_{SL}$} even further. If we aim to cover all the records in the test set, the accuracy of S-GCN reaches $0.85$, whereas that of RLS reaches  $0.78$.
Moreover, setting $k = 2$ allows us to raise the accuracy to \new{$0.92$} with a high coverage of $0.87$.
On the other hand, RLS can only reach a comparable accuracy of $0.90$ by dropping coverage to $0.60$.

The performance of our model is much better on overlap-all, as this dataset is more suitable for learning.
We observe a similar pattern for short names as their usage improves the accuracy for both systems.
However, coverage for RLS drops faster for comparable improvements in accuracy.
Most notably, \new{S-GCN$_{SL}$ can achieve an almost perfect accuracy of $0.98$ with a high coverage of $0.93$}.
Thus, we can set the parameters of the system such that we are highly confident that predictions are correct, while obtaining a match in more than $90\%$ of the cases. On the other hand, reaching a very high accuracy of $0.98$ with RLS requires dropping coverage to $0.42$.

\subsection{Learning Company Embeddings}
\label{sec:embeddings}

In this section, we study the consistency and the quality of learned S-GCN embeddings. We employ \textit{silhouette scores}~\cite{Rousseeuw1987} as a measure of how well the entities of company families are clustered in the S-GCN embedding space. A silhouette value of an entity measures how close a company entity is to its family in the projection. The silhouette value is defined in a range of [$-1$, 1], where a high score indicates that the entity is well coordinated with its company family. If the average silhouette score (over all entities) has a positive value, then families are disjoint on average, whereas if the score is negative, families are overlapping in the embedding space. We computed the silhouette score for S-GCN \new{where the features are initialized with the embeddings from the BERT model} and the \new{pre-trained} BERT model on the two datasets overlap-1 and overlap-all. The results are shown in Table \ref{tab:sscore}. Average silhouette scores for S-GCN are always positive and \new{are higher that the results of the sole BERT model demonstrating that S-GSN is able to produce more discriminating features by using additional information flow of the company graph.} 

\begin{figure}[!htb]
	\begin{center}
		\includegraphics[width=0.55\linewidth, trim=48 48 0 0, clip]{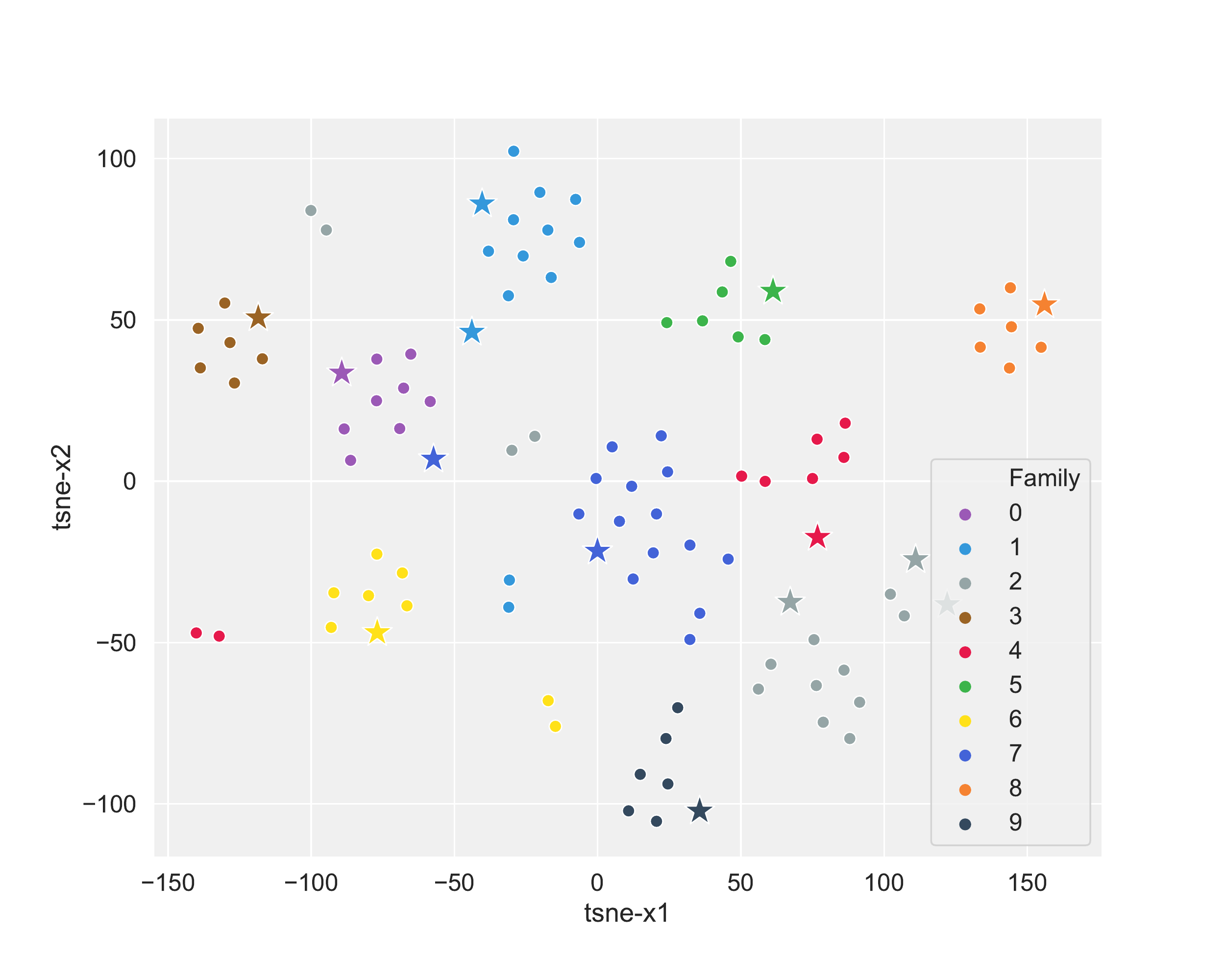}  
		\caption{2D representation of 10 randomly selected company families computed by S-GCN, where dots represent  records $r \in R$, while stars refer to records $q \in Q$}
		\label{fig:s2dClusters}
	\end{center}
\end{figure}

To demonstrate the results visually, we randomly selected ten company families from the  embeddings produced by our S-GCN model (on overlap-all with short names) and projected 300 embedding dimensions into 2D with $t$-SNE~\cite{Maaten2008}. Figure~\ref{fig:s2dClusters} shows the 2D projection of the data, where the dots correspond to reference records from $R$ (training examples), colors indicate company families, and stars represent corresponding records from $Q$ (test records). We see that $q$ records from the database $Q$ (stars) can be matched to their families even in 2D space.

\begin{table}[!htb]
\caption{Average silhouette score of company families}
\label{tab:sscore}
\centering
\begin{tabular}{|l|c|c|}
\hline
\multirow{2}{*}{} & \multicolumn{2}{l|}{\textbf{Avg. silhouette score}}  \\ \cline{2-3} 
 & \textbf{\new{BERT +} S-GCN } & \textbf{\new{only} BERT} \\ \hline
 overlap-1 & 0.22 & $-0.12$   \\ \hline
overlap-all & 0.63 & 0.01 \\ \hline
\end{tabular}%
\end{table}

 \new{
\section{Discussion}}
\label{sec:challenges}
Although the experimental results are promising, we envision several challenges that need to be addressed for a proper application of GNNs to data integration tasks. This section lists some of the main problems that have to be investigated and require further research.

\parg{Graph extraction}
Generating a graph-structured representation of the data is a crucial and time-consuming process, with a potentially strong impact on subsequent design choices as well as on the final performance of the model.
Relational databases usually provide many different connections between entities, and can typically be modeled as a graph in several different ways. 
Defining exactly what kind of entities should be represented as a node and what kind of relations should be included as edges strongly affects the way information is propagated between the neighboring nodes during training.

Moreover, as noted in~\cite{Xirogiannopoulos2017}, extracting a graph from a relational database has several other potential applications, thanks to the insights that can be drawn from explicit modeling of entities and their connections.
Some automatic or semi-automatic approaches have been proposed in the literature.
For instance, a domain-specific language (DSL) based on Datalog is used in~\cite{Xirogiannopoulos2017} to specify graph-extraction queries.
Recently, an approach to  cleaning raw tabular data automatically and generating linked data has been introduced in~\cite{Sukhobok2016}.
Additionally, \cite{Arenas2012}~describes a direct mapping from the data in a relational database to a knowledge graph. The algorithms defined in~\cite{Arenas2012} can even be applied to multiple tables, where foreign keys are used to establish a reference from a given row in a table to exactly one row in a potentially different table.

In spite of these results, the process of extracting graph representations of relational databases is still challenging and labor intensive, and is often performed manually.
The need for automatic approaches is an even stronger requirement for unstructured data. Some recent work in this area includes~\cite{Gangemi2016} and~\cite{Das2019}.

\parg{Handling different relations}
As previously discussed in Section~\ref{sec:graph_extraction}, in our application we only use a single relation type to connect both aliases of the same company and companies belonging to the same family.
More precisely, in our use case, an edge in the graph $R^*$ is a pair of the type $(r^*_i, r^*_j)$, where $r^*_i$ and $r^*_j$ are adjacent nodes in the graph. 
However, in several realistic large-scale knowledge bases, this assumption is usually restrictive.
Relations between two entities in a KG can be of type $\tau \in \mathcal{T}$, where $\mathcal{T}$ is the set of all  possible relations between two entities in the graph.
In this case, edges are not represented as pairs, but are modeled as triples of the form $(r_i, \tau, r_j)$.
This poses some challenges related to processing large-scale relational data with GNNs.
Recently, a variation of GCN that is capable of making relation-specific transformations based on the type and direction of an edge was introduced in~\cite{Schlichtkrull2018}.
The model, termed Relational GCN (R-GCN), is a special case of the message-passing neural network proposed in~\cite{Gilmer2017}, as it propagates messages while taking into account relation types by learning a different weight matrix for each $\tau \in \mathcal{T}$.
In practice, this poses several challenges when dealing with highly multi-relational data.
Indeed, the number of parameters in the model grows with the number of relations in the graph.
This can result in very large models, or even lead to overfitting on rare relations.
Some regularization strategies to address this issue have been introduced in~\cite{Schlichtkrull2018}.
These techniques allow  the total number of parameters to be limited, either by sharing parameters among weight matrices or enforcing sparsity.

\parg{Data augmentation}
Some nodes in $R^*$ or $Q^*$ might have a small degree of incoming edges. Artificially generating aliases for a node with a low degree and connecting them to the original node might lead to a better generalization ability of GNNs and consequently improve its accuracy. Information from aliases propagates towards the original node, where it accumulates with the node information resulting in the output embedding. In our use case, we improved the results greatly by generating short company names from their original names (see Section \ref{sec:short_names}). However, generating aliases can be labor-intensive and might require manual tuning.


\section{Conclusion}
\label{sec:conclusion}

We have investigated how graph modeling and graph neural networks can be leveraged for data integration. 
Experimental results show that GNNs and specifically S-GCNs are effective in performing record linkage. \new{The key lessons learned from our work are the following: \textit{(i)} it is beneficial to model the graph structure for the record linkage task as it allows a natural way to transfer information among the nodes, and \textit{(ii)} graph neural networks can improve data integration with structured and unstructured sources by allowing high automation with close to none labelling costs.}
Future work will aim at exploring several aspects that are both methodological and architectural.  On the methods side, we aim at providing guidance for cases where attributes in a record are better represented as node features rather than separate nodes.
In addition, we will explore different training strategies and  GNN architectures.
We hope that our results will encourage other \new{data integration researchers and practitioners to contribute both empirically and theoretically on graph-based data integration challenges.}





\balance

\begin{thebibliography}{10}

\bibitem{Arenas2012}
M.~Arenas, A.~Bertails, E.~Prud'hommeaux, and J.~Sequeda.
\newblock {A Direct Mapping of Relational Data to RDF}.
\newblock \url{https://www.w3.org/TR/rdb-direct-mapping/}, September 2012.

\bibitem{Azmy2019}
M.~Azmy, P.~Shi, J.~Lin, and I.~F. Ilyas.
\newblock Matching entities across different knowledge graphs with graph
  embeddings.
\newblock {\em CoRR}, abs/1903.06607, 2019.

\bibitem{Battaglia2018}
P.~W. Battaglia, J.~B. Hamrick, V.~Bapst, A.~Sanchez{-}Gonzalez, V.~F.
  Zambaldi, M.~Malinowski, A.~Tacchetti, D.~Raposo, A.~Santoro, R.~Faulkner,
  {\c{C}}.~G{\"{u}}l{\c{c}}ehre, F.~Song, A.~J. Ballard, J.~Gilmer, G.~E. Dahl,
  A.~Vaswani, K.~Allen, C.~Nash, V.~Langston, C.~Dyer, N.~Heess, D.~Wierstra,
  P.~Kohli, M.~Botvinick, O.~Vinyals, Y.~Li, and R.~Pascanu.
\newblock Relational inductive biases, deep learning, and graph networks.
\newblock {\em CoRR}, abs/1806.01261, 2018.

\bibitem{Bromley1993}
J.~Bromley, I.~Guyon, Y.~LeCun, E.~S\"{a}ckinger, and R.~Shah.
\newblock Signature verification using a "siamese" time delay neural network.
\newblock NIPS'93, pages 737--744, 1993.

\bibitem{Chopra2005}
S.~Chopra, R.~Hadsell, and Y.~LeCun.
\newblock Learning a similarity metric discriminatively, with application to
  face verification.
\newblock CVPR '05, pages 539--546, 2005.

\bibitem{cybenko1989}
G.~Cybenko.
\newblock Approximation by superpositions of a sigmoidal function.
\newblock {\em Mathematics of control, signals and systems}, 2(4):303--314,
  1989.

\bibitem{Das2018}
R.~Das, T.~Munkhdalai, X.~Yuan, A.~Trischler, and A.~McCallum.
\newblock Building dynamic knowledge graphs from text using machine reading
  comprehension.
\newblock {\em arXiv preprint arXiv:1810.05682}, 2018.

\bibitem{Das2019}
R.~Das, T.~Munkhdalai, X.~Yuan, A.~Trischler, and A.~McCallum.
\newblock Building dynamic knowledge graphs from text using machine reading
  comprehension.
\newblock In {\em {ICLR}}, 2019.

\bibitem{Devlin2019}
J.~Devlin, M.~Chang, K.~Lee, and K.~Toutanova.
\newblock {BERT:} pre-training of deep bidirectional transformers for language
  understanding.
\newblock In {\em Proceedings of {NAACL-HLT}}, pages 4171--4186, 2019.

\bibitem{Dong2018}
X.~L. Dong and T.~Rekatsinas.
\newblock Data integration and machine learning: A natural synergy.
\newblock SIGMOD '18, pages 1645--1650, 2018.

\bibitem{Ganea2017}
O.~Ganea and T.~Hofmann.
\newblock Deep joint entity disambiguation with local neural attention.
\newblock In {\em Proceedings of the 2017 Conference on Empirical Methods in
  Natural Language Processing, {EMNLP} 2017, Copenhagen, Denmark, September
  9-11, 2017}, pages 2619--2629, 2017.

\bibitem{Gangemi2016}
A.~Gangemi, V.~Presutti, D.~Reforgiato~Recupero, A.~Nuzzolese, F.~Draicchio,
  and M.~Mongiovi.
\newblock Semantic web machine reading with fred.
\newblock {\em Semantic Web}, 8:1--21, 09 2016.

\bibitem{Getoor2012}
L.~Getoor and A.~Machanavajjhala.
\newblock Entity resolution: Theory, practice \& open challenges.
\newblock {\em PVLDB}, 2012.

\bibitem{Gilmer2017}
J.~Gilmer, S.~S. Schoenholz, P.~F. Riley, O.~Vinyals, and G.~E. Dahl.
\newblock Neural message passing for quantum chemistry.
\newblock ICML'17, 2017.

\bibitem{Gori2005}
M.~{Gori}, G.~{Monfardini}, and F.~{Scarselli}.
\newblock A new model for learning in graph domains.
\newblock In {\em Proceedings of the IEEE International Joint Conference on
  Neural Networks}, volume~2, pages 729--734 vol. 2, July 2005.

\bibitem{Gottapu2016}
R.~D. Gottapu, C.~Dagli, and B.~Ali.
\newblock Entity resolution using convolutional neural network.
\newblock {\em Procedia Computer Science}, 95:153 -- 158, 2016.
\newblock Complex Adaptive Systems Los Angeles, CA November 2-4, 2016.

\bibitem{gschwind2019}
T.~Gschwind, C.~Miksovic, J.~Minder, K.~Mirylenka, and P.~Scotton.
\newblock Fast record linkage for company entities.
\newblock {\em IEEE Big Data 2019, December 9-12, 2019, Los Angeles, CA, USA},
  2019.

\bibitem{Hadsell2006}
R.~Hadsell, S.~Chopra, and Y.~LeCun.
\newblock Dimensionality reduction by learning an invariant mapping.
\newblock CVPR '06, 2006.

\bibitem{Guyon2017}
W.~Hamilton, Z.~Ying, and J.~Leskovec.
\newblock Inductive representation learning on large graphs.
\newblock In I.~Guyon, U.~V. Luxburg, S.~Bengio, H.~Wallach, R.~Fergus,
  S.~Vishwanathan, and R.~Garnett, editors, {\em Advances in Neural Information
  Processing Systems 30}, pages 1024--1034. Curran Associates, Inc., 2017.

\bibitem{Khalil2017}
E.~B. Khalil, H.~Dai, Y.~Zhang, B.~Dilkina, and L.~Song.
\newblock Learning combinatorial optimization algorithms over graphs.
\newblock In I.~Guyon, U.~von Luxburg, S.~Bengio, H.~M. Wallach, R.~Fergus,
  S.~V.~N. Vishwanathan, and R.~Garnett, editors, {\em NIPS 2017}, 2017.

\bibitem{Kipf2017}
T.~N. Kipf and M.~Welling.
\newblock Semi-supervised classification with graph convolutional networks.
\newblock In {\em {ICLR} 2017}, 2017.

\bibitem{Kool2019}
W.~Kool, H.~van Hoof, and M.~Welling.
\newblock Attention, learn to solve routing problems!
\newblock In {\em International Conference on Learning Representations}, 2019.

\bibitem{Ktena2017}
S.~I. Ktena, S.~Parisot, E.~Ferrante, M.~Rajchl, M.~C.~H. Lee, B.~Glocker, and
  D.~Rueckert.
\newblock Distance metric learning using graph convolutional networks:
  Application to functional brain networks.
\newblock {\em CoRR}, abs/1703.02161, 2017.

\bibitem{Lehmann2015}
J.~Lehmann, R.~Isele, M.~Jakob, A.~Jentzsch, D.~Kontokostas, P.~N. Mendes,
  S.~Hellmann, M.~Morsey, P.~van Kleef, S.~Auer, and C.~Bizer.
\newblock Dbpedia - {A} large-scale, multilingual knowledge base extracted from
  wikipedia.
\newblock {\em Semantic Web}, 6(2):167--195, 2015.

\bibitem{Lin2018}
H.~Lin and S.~Jegelka.
\newblock Resnet with one-neuron hidden layers is a universal approximator.
\newblock In {\em NeurIPS 2018}, 2018.

\bibitem{Liu2018}
B.~Liu, T.~Zhang, D.~Niu, J.~Lin, K.~Lai, and Y.~Xu.
\newblock Matching long text documents via graph convolutional networks.
\newblock 02 2018.

\bibitem{Logeswaran2019}
L.~Logeswaran, M.~Chang, K.~Lee, K.~Toutanova, J.~Devlin, and H.~Lee.
\newblock Zero-shot entity linking by reading entity descriptions.
\newblock {\em CoRR}, abs/1906.07348, 2019.

\bibitem{Loster2017}
M.~Loster, Z.~Zuo, F.~Naumann, O.~Maspfuhl, and D.~Thomas.
\newblock Improving company recognition from unstructured text by using
  dictionaries.
\newblock In {\em EDBT 2017}, pages 610--619, 2017.

\bibitem{Ma2018}
G.~Ma, N.~K. Ahmed, T.~L. Willke, D.~Sengupta, M.~W. Cole, N.~B. Turk{-}Browne,
  and P.~S. Yu.
\newblock Similarity learning with higher-order proximity for brain network
  analysis.
\newblock {\em CoRR}, abs/1811.02662, 2018.

\bibitem{Mikolov2013}
T.~Mikolov, I.~Sutskever, K.~Chen, G.~S. Corrado, and J.~Dean.
\newblock Distributed representations of words and phrases and their
  compositionality.
\newblock In C.~J.~C. Burges, L.~Bottou, Z.~Ghahramani, and K.~Q. Weinberger,
  editors, {\em NIPS 2013}, pages 3111--3119, 2013.

\bibitem{mirylenka2016applicability}
K.~Mirylenka, C.~Miksovic, and P.~Scotton.
\newblock Applicability of latent dirichlet allocation for company modeling.
\newblock In {\em Industrial Conference on Data Mining (ICDM'2016)}, 2016.

\bibitem{mirylenka2016recurrent}
K.~Mirylenka, C.~Miksovic, and P.~Scotton.
\newblock Recurrent neural networks for modeling company-product time series.
\newblock {\em Proceedings of the workshop on Advanced Analytics and Learning
  on Temporal Data (AALTD) in conjunction with ECML PKDD}, pages 29--36, 2016.

\bibitem{mirylenka2019similarity}
K.~Mirylenka, P.~Scotton, C.~A.~M. Czasch, and A.~Schade.
\newblock Similarity matching systems and methods for record linkage, Nov.~21
  2019.
\newblock US Patent App. 15/980,066.

\bibitem{mirylenka2019linking}
K.~Mirylenka, P.~Scotton, C.~Miksovic, and S.-E. Bariol~Alaoui.
\newblock Linking it product records.
\newblock In {\em Data Integration and Applications Workshop (DINA) of ECML
  PKDD}, 2019.

\bibitem{mirylenka2019hidden}
K.~Mirylenka, P.~Scotton, C.~Miksovic, and J.~Dillon.
\newblock Hidden layer models for company representations and product
  recommendations.
\newblock In {\em EDBT}, pages 468--476, 2019.

\bibitem{Monti2017}
F.~Monti, D.~Boscaini, J.~Masci, E.~Rodol{\`{a}}, J.~Svoboda, and M.~M.
  Bronstein.
\newblock Geometric deep learning on graphs and manifolds using mixture model
  cnns.
\newblock In {\em 2017 {IEEE} Conference on Computer Vision and Pattern
  Recognition, {CVPR} 2017, Honolulu, HI, USA, July 21-26, 2017}, pages
  5425--5434, 2017.

\bibitem{Mudgal2018}
S.~Mudgal, H.~Li, T.~Rekatsinas, A.~Doan, Y.~Park, G.~Krishnan, R.~Deep,
  E.~Arcaute, and V.~Raghavendra.
\newblock Deep learning for entity matching: A design space exploration.
\newblock In {\em SIGMOD '18}, pages 19--34, 2018.

\bibitem{pennington2014}
J.~Pennington, R.~Socher, and C.~D. Manning.
\newblock Glove: Global vectors for word representation.
\newblock In {\em EMNLP}, 2014.

\bibitem{pershina2015}
M.~Pershina, M.~Yakout, and K.~Chakrabarti.
\newblock Holistic entity matching across knowledge graphs.
\newblock In {\em 2015 IEEE International Conference on Big Data (Big Data)},
  pages 1585--1590. IEEE, 2015.

\bibitem{Rousseeuw1987}
P.~Rousseeuw.
\newblock Silhouettes: a graphical aid to the interpretation and validation of
  cluster analysis.
\newblock 1987.

\bibitem{Scarselli2009}
F.~Scarselli, M.~Gori, A.~C. Tsoi, M.~Hagenbuchner, and G.~Monfardini.
\newblock The graph neural network model.
\newblock {\em {IEEE} Trans. Neural Networks}, 20(1):61--80, 2009.

\bibitem{Schlichtkrull2018}
M.~S. Schlichtkrull, T.~N. Kipf, P.~Bloem, R.~van~den Berg, I.~Titov, and
  M.~Welling.
\newblock Modeling relational data with graph convolutional networks.
\newblock In A.~Gangemi, R.~Navigli, M.~Vidal, P.~Hitzler, R.~Troncy,
  L.~Hollink, A.~Tordai, and M.~Alam, editors, {\em {ESWC} 2018}, 2018.

\bibitem{Schroff2015}
F.~Schroff, D.~Kalenichenko, and J.~Philbin.
\newblock Facenet: A unified embedding for face recognition and clustering.
\newblock {\em CVPR}, pages 815--823, 2015.

\bibitem{Selsam2018}
D.~Selsam, M.~Lamm, B.~B{\"{u}}nz, P.~Liang, L.~de~Moura, and D.~L. Dill.
\newblock Learning a {SAT} solver from single-bit supervision.
\newblock {\em CoRR}, abs/1802.03685, 2018.

\bibitem{Shervashidze2011}
N.~Shervashidze, P.~Schweitzer, E.~J. van Leeuwen, K.~Mehlhorn, and K.~M.
  Borgwardt.
\newblock Weisfeiler-lehman graph kernels.
\newblock {\em J. Mach. Learn. Res.}, 12:2539--2561, 2011.

\bibitem{Fakhraei2019}
J.~L.~A. Shobeir~Fakhraei, Joel~Mathew.
\newblock Nseen: Neural semantic embedding for entity normalization.
\newblock {\em PKDD}, 2019.

\bibitem{Sukhobok2016}
D.~Sukhobok, N.~Nikolov, A.~Pultier, X.~Ye, A.~J. Berre, R.~Moynihan,
  B.~Roberts, B.~Elves{\ae}ter, M.~Nivethika, and D.~Roman.
\newblock Tabular data cleaning and linked data generation with grafterizer.
\newblock In H.~Sack, G.~Rizzo, N.~Steinmetz, D.~Mladenic, S.~Auer, and
  C.~Lange, editors, {\em The Semantic Web - {ESWC} 2016 Satellite Events,
  Revised Selected Papers}, pages 134--139, 2016.

\bibitem{Trivedi2018}
R.~Trivedi, B.~Sisman, X.~L. Dong, C.~Faloutsos, J.~Ma, and H.~Zha.
\newblock Linknbed: Multi-graph representation learning with entity linkage.
\newblock In {\em Proceedings of the 56th Annual Meeting of the Association for
  Computational Linguistics, {ACL} 2018, Melbourne, Australia, July 15-20,
  2018, Volume 1: Long Papers}, pages 252--262, 2018.

\bibitem{Maaten2008}
L.~van~der Maaten and G.~Hinton.
\newblock Visualizing data using t-sne, 2008.

\bibitem{Vaswani2017}
A.~Vaswani, N.~Shazeer, N.~Parmar, J.~Uszkoreit, L.~Jones, A.~N. Gomez,
  L.~Kaiser, and I.~Polosukhin.
\newblock Attention is all you need.
\newblock In {\em Advances in Neural Information Processing Systems 30}, pages
  6000--6010, 2017.

\bibitem{Velickovic2018}
P.~Velickovic, G.~Cucurull, A.~Casanova, A.~Romero, P.~Li{\`{o}}, and
  Y.~Bengio.
\newblock Graph attention networks.
\newblock In {\em {ICLR}}, 2018.

\bibitem{Wang2017H}
C.~Wang, X.~Zhang, and X.~Lan.
\newblock How to train triplet networks with 100k identities?
\newblock {\em 2017 IEEE International Conference on Computer Vision Workshops
  (ICCVW)}, pages 1907--1915, 2017.

\bibitem{Wang2018}
X.~Wang, R.~B. Girshick, A.~Gupta, and K.~He.
\newblock Non-local neural networks.
\newblock In {\em IEEE {CVPR} 2018}, pages 7794--7803, 2018.

\bibitem{weisfeiler1968}
B.~Weisfeiler and A.~A. Lehman.
\newblock A reduction of a graph to a canonical form and an algebra arising
  during this reduction.
\newblock {\em Nauchno-Technicheskaya Informatsia}, 2(9):12--16, 1968.

\bibitem{Xirogiannopoulos2017}
K.~Xirogiannopoulos and A.~Deshpande.
\newblock Extracting and analyzing hidden graphs from relational databases.
\newblock In {\em {SIGMOD}}, 2017.

\bibitem{Xu2019}
K.~Xu, W.~Hu, J.~Leskovec, and S.~Jegelka.
\newblock How powerful are graph neural networks?
\newblock {\em ICLR 2019}, 2019.

\bibitem{Xu2018}
K.~Xu, C.~Li, Y.~Tian, T.~Sonobe, K.~Kawarabayashi, and S.~Jegelka.
\newblock Representation learning on graphs with jumping knowledge networks.
\newblock In {\em Proceedings of {ICML} 2018}, 2018.

\bibitem{Ying2018}
Z.~Ying, J.~You, C.~Morris, X.~Ren, W.~L. Hamilton, and J.~Leskovec.
\newblock Hierarchical graph representation learning with differentiable
  pooling.
\newblock In {\em NeurIPS 2018}, pages 4805--4815, 2018.

\bibitem{Jiaxuan2018NIPS}
J.~You, B.~Liu, Z.~Ying, V.~Pande, and J.~Leskovec.
\newblock Graph convolutional policy network for goal-directed molecular graph
  generation.
\newblock In S.~Bengio, H.~Wallach, H.~Larochelle, K.~Grauman, N.~Cesa-Bianchi,
  and R.~Garnett, editors, {\em NIPS 18}. 2018.

\bibitem{Jiaxuan2018}
J.~You, R.~Ying, X.~Ren, W.~L. Hamilton, and J.~Leskovec.
\newblock Graphrnn: {A} deep generative model for graphs.
\newblock {\em CoRR}, abs/1802.08773, 2018.

\bibitem{Zhang2018}
M.~Zhang, Z.~Cui, M.~Neumann, and Y.~Chen.
\newblock An end-to-end deep learning architecture for graph classification.
\newblock In {\em Proceedings of AAAI-18, IAAI-18, and EAAI-18}, 2018.

\bibitem{Zhang2019}
W.~Zhang, K.~Shu, H.~Liu, and Y.~Wang.
\newblock Graph neural networks for user identity linkage.
\newblock {\em CoRR}, abs/1903.02174, 2019.

\end{thebibliography}

\end{document}